\documentclass[structabstract]{aa}  
\usepackage{natbib}
\usepackage{graphicx}
\usepackage{txfonts}
\usepackage{rotating}
\usepackage{subfigure}
\newcommand{\ha}[0]{H$\alpha$ }
\newcommand{\galex}[0]{\textit{GALEX} }
\newcommand{\spitzer}[0]{\textit{Spitzer} }
\newcommand{\micron}[0]{$\mu${\rm m}}
\newcommand{\starburst}[0]{\textit{Starburst99} }
\def\hi{\relax \ifmmode {\mbox H\,{\scshape i}}\else H\,{\scshape i}\fi}
\def\hii{\relax \ifmmode {\mbox H\,{\scshape ii}}\else H\,{\scshape ii}\fi}
\def\nii{\relax \ifmmode {\mbox N\,{\scshape ii}}\else N\,{\scshape ii}\fi}

\begin{document}
  \title{Multiwavelength study of the star formation in the bar of NGC 2903}

   \author{G. Popping
          \inst{1}
          \and
          I. P\'erez \inst{2,3}
          \and
          A. Zurita \inst{2,3}
          }

   \institute{Kapteyn Astronomical Institute, PO Box 800, 9700 AV Groningen, The Netherlands\\
              \email{G.Popping@astro.rug.nl}
         \and
             Dep. F\'isica Te\'orica y del Cosmos, Campus de Fuentenueva, Universidad de Granada, 18071 Granada, Spain \and Instituto Carlos I de F\'isica Te\'orica y Computaci\'on, Spain 
               }

   \date{}
\abstract
   {} 
    {NGC~2903 is a nearby barred spiral with an active starburst in the center and \hii\ regions distributed along its bar. We aim to analyse the star formation properties in the bar region of NGC~2903 and study the links with the typical bar morphological features.}
   {A combination of space and ground--based data from the far--ultraviolet to the sub--millimeter spectral ranges is used to create a
    panchromatic view of the NGC~2903 bar.
    We produce two catalogues: one for the current star formation regions, as traced by the \ha compact emission, and 
    a second one for the ultraviolet (UV) emitting knots, containing positions and luminosities. From them we have obtained 
    ultraviolet colours, star formation rates, dust attenuation and \ha EWs, and their spatial distribution have been 
    analysed. 
    Stellar cluster ages have been estimated using stellar population synthesis models ({\em Starburst99}).}
    {NGC~2903 is a  complex galaxy, with a very different morphology on each spectral band. The CO~($J$=1-0) and the 3.6~\micron\ emission trace each other in a clear barred structure, while the \ha leads both components and it has an s--shape distribution. The UV emission is patchy and does not resemble a bar. 
The UV emission is also characterised by a number of regions located symmetrically with respect to the galaxy center, almost perpendicular to the bar,  in a spiral shape covering the inner $\sim2.5$kpc. These regions do not show a significant \ha nor 24~\micron\ emission. We have estimated ages for these regions ranging from 150 to 320~Myr, being older than the rest  of the UV knots, which have ages lower than 10~Myr. 
The SFR calculated from the UV emission is  $\sim$0.4~M$_{\odot}$~yr$^{-1}$, compatible with the SFR as derived from \ha calibrations ($\sim$1~M$_{\odot}$~yr$^{-1}$).}
{}
 \maketitle

\begin{table}[b]
\caption{Log of  ground--based \ha observations.  \label{tab:log}}
\begin{tabular}{ccccc}
\hline
Band & Date & Filter & Exp. Time & Seeing\\
\hline
H$\alpha$ & 29 Oct 2007 & WFCH6568 & 3x1200s & 1.4"\\
R & 30 Oct 2007 & HARRIS R & 3x250s & 2.1"\\
\hline
\end{tabular}
\end{table}

%
\section{Introduction}

In the years after the launch of the \textit{Spitzer Space Telescope} and the \textit{Galaxy Evolution Explorer} ({\em galex}) knowledge on the star formation (SF) in galaxies have grown considerably. The panchromatic view of nearby galaxies offered by large surveys carried out from these space telescope missions gives an extraordinary database to study star formation in galaxies. In particular, it allows us to link general galaxy properties with the the local interstellar medium (ISM) properties and the galaxy dynamics. These links are crucial for understanding SF in galaxies. In nearby galaxies these sets of multi-wavelength data give the opportunity to study in detail, with good spatial resolution, the location and properties of relatively young populations, recent massive star formation and dust attenuation (e.g. Calzetti et al. 2005; Tamura et al. 2009; Rela\~no \& Kennicutt 2009).\nocite{tamura}

New dust and SF indicators have been developed and calibrated using all the available bands. The combination of 24~\micron\ and \ha emission has become a reliable indicator of dust attenuation (Kennicutt et al. 2007; Rela\~no \& Kennicutt 2009)\nocite{Kennicutt2007,Relano2009}. Both bands are linked to star formation processes, with the \ha emission originating from  the recombination of hydrogen  in the surrounding medium of very recently formed (less than a few Myr)  massive stars, and the 24 \micron\ emission as tracing local star formation radiation obscured by dust (Calzetti et al. 2005). Extinction corrected UV emission can also be used to retrieve the star formation rate (SFR) (Kennicutt 1998)\nocite{Kennicutt1998} and to get some insight into the star formation occurred in last  Gyr (e.g. Bianchi et al. 2005; Hibbard et al. 2005). This feature makes a combination of UV,  \ha and  24 \micron\ emission ideal to reconstruct the recent star formation history in a galaxy by studying the location and properties of individual clusters and those of the gas and dust emission around them.\nocite{bianchi2005,hibbard2005}\\

To this date, and to our knowledge, only a few galaxies have  been analysed in this way;  M51 (Calzetti et al. 2005), M81 (P\'erez-Gonz\'alez et al. 2006); M33 (Rela\~no \& Kennicutt 2009; Verley et al. 2009); NGC~7331 (Thilker et al. 2007).\nocite{perezgonzalez,verley2009,thilker2007}

Barred galaxies offer a useful tool to investigate the physical conditions that favour star formation in galaxies. The motions in bars are characterised by non-circular motions that push the gas into intersecting orbits where shocks and star formation can be triggered. The position and strength of the shocks are determined by the bar potential, the global dynamics within the bar region is driven by the bar. We have now a relatively good understanding of the gas behaviour under a bar potential (e.g. P\'erez et al. 2004)\nocite{perez2004}  and this knowledge can be used to understand the conditions triggering star formation. Bars and their surroundings host extreme physical conditions and a variety of ISM environments. They are perfect places to study the link between the conditions favouring star formation with the galaxy dynamics. Therefore,  a panchromatic view revealing the history of star formation in bars can give a unique insight into the links of star formation and the galaxy dynamics, which could also help  to understand how bars form and evolve.  

This work presents a detailed multi-wavelength study of the star formation in the bar region of NGC~2903 analysing the correlations between the location and ages of the {\it young} stellar clusters with the morphology of the bar.  This is done by analysing the emission in \ha, UV and 24 \micron\ as well as optical data from the Sloan Digital Sky Survey (SDSS), and using 8 \micron\ and CO~(J$=$1-0) emission as complementary data.

NGC 2903 is chosen for this research for a number of reasons: it is  close by (8.9 Mpc; Drozdovsky \& Karachentsev 2000)  \nocite{Drozdovsky2000}allowing us to have high spatial resolution ($\sim$~43~pc~arcsec$^{-1}$), and  it is isolated from large companions, preventing major merger effects in the results. Recent work (Irwin et al. 2009) on the \hi\ content of NGC~2903 has shown that it possesses a large \hi\ envelope of around three times its optical size. They also found a small \hi\ companion 64~kpc away from the galaxy in projection,  which adds to a previously known small stellar companion. No clear sign of interaction has been found so far.\nocite{irwin2009}

NGC~2903 is a SBd galaxy showing a symmetric strong bar considered typical for this class of galaxies (Laurikainen \& Salo 2002).\nocite{Laurikainen2002} Previous observations have shown large amounts  of \ha emission along the bar and not only  at the ends of the bar and nuclear region (Sheth et al. 2002).  The CO($J=$1 - 0), \spitzer and \galex data available makes this galaxy an ideal object for a multi wavelength study to retrieve insight in star formation history in bars.

A previous study by Leon et al. (2008)\nocite{Leon2008} on the NGC 2903 bar  showed that HCN(1-0) is distributed along the bar and in the center. They compared the star formation rate ratio between the bar and the center with results from numerical simulations by Martin \& Friedli (1997)\nocite{MartinFriedli1997}. This made them propose that the bar in NGC 2903 has an age between 200 and 600 Myr.

The plan for the article is the following: in section \ref{ObservationsDataReduction} we present the observational data; in section~\ref{morphology} we analyse the general morphology of NGC~2903. We  present our methodology to obtain the bar \hii\ regions and UV emission knots catalogues in section~\ref{photometry}. 
The following section contains the main results regarding EW$_{H\alpha}$, star formation rates, UV colours and ages of the stellar clusters.  In section~\ref{sec:discussion} we discuss our main results and finally, we end  with a summary and conclusions.


\section{Observations and data reduction}
\label{ObservationsDataReduction}
\subsection{H$\alpha$ imaging}
\label{ha}

The H$\alpha$ data of NGC~2903 was obtained with the Wide Field Camera at the  2.5m Isaac Newton Telescope at the Roque de
los Muchachos Observatory in October 2007. A summary of the observations is presented in Table 1. The galaxy was observed through a 95 \AA{} width narrowband filter with a total exposure time of 3600s, and an R--band filter, used for the continuum subtraction, with an exposure time of 750s. 
We carried out the overscan subtraction, bias and flatfield correction using standard reduction tasks within IRAF\footnote{IRAF is distributed by the National Optical Astronomy Observatories, which is operated by the Association of Universities for Research in Astronomy, Inc. (AURA) under cooperative agreement with the National Science Foundation.}. The sky subtraction was carried out by fitting a first order polynomial to a sky map created using the mean sky values derived from several positions on the image, free from foreground stars and galaxy emission. The H$\alpha$+continuum and
R--band images were then aligned using positions of field stars in the images and combined separately to produce the final images.
Before continuum subtraction the  H$\alpha$+continuum image was degraded to the resolution of the R--band image ($\sim2.1$'').

The continuum subtraction was done as described by \cite{Relano2005}.
We used 11 non--saturated foreground stars  in both the \ha
(ON--band) and in the continuum R--band (OFF--band)
images. The fluxes of these stars yielded a mean flux ratio
ON--band/OFF--band of 0.34$\pm$0.02. This value is then used as a
starting point scaling factor to generate a set of OFF--band images 
scaled with factors ranging from 0.32 to 0.36. These images were
afterwards subtracted from the ON--band image to produce images free of
continuum emission. After a close inspection of the resulting images we
adopted a value of 0.34$\pm$0.01 for the continuum scaling factor. The
uncertainty of the continuum scaling factor produces differences of up
to $\sim$3\% in the integrated fluxes of the \hii\ regions.\nocite{kennicutt1983}

\begin{figure}
\includegraphics[scale = 0.5]{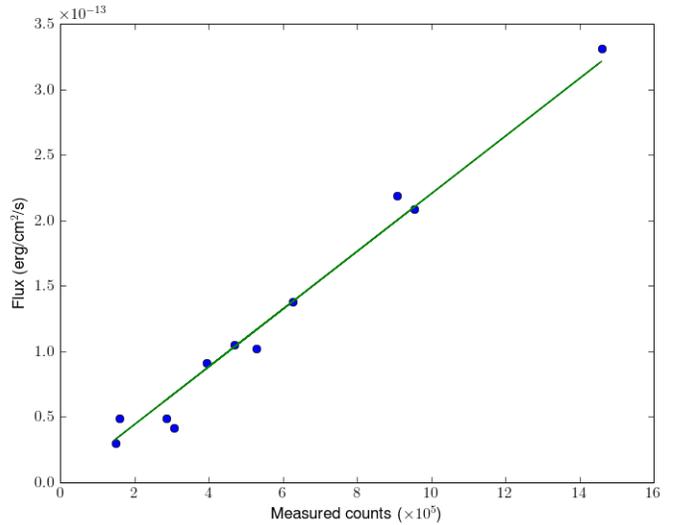}
\caption{Scaling relation from our measured H$\alpha$+[\nii] fluxes in
counts to earlier measurements made by Mayya et al. (1994) for the same
\hii\ regions. The \hii\ region fluxes measured by Mayya et
al. (1994) also contain the [\nii] emission.\label{fig:ha_scaling}}
\end{figure}

The astrometry on the \ha image was performed using the USNO2 catalogue coordinates for the foreground stars of the galaxy images, resulting in an accuracy of $\sim$0.39''. The observations were made during non photometric conditions. Therefore, for the flux calibration, we used previously reported fluxes of 11 regions \hii\ regions located in the disk of NGC~2903 published by \cite{Mayya1994}.  We compared  the flux reported there with our measured flux (in ADU) using identical apertures. We then calculated the ratio for each region and took the median as our final H$\alpha$ flux calibration factor ((2.2$\pm$0.4)$\times10^{-19}$erg~s$^{-1}$~cm$^{-2}$~count$^{-1}$). The error was calculated by taking the standard deviation of the flux ratio for all the stars. The scaling relation is presented in Fig.~\ref{fig:ha_scaling}. The 2$\sigma$ sensitivity limit of the our final \ha image is  $6.2 \times10^{-17}$ erg s$^{-1}$ cm$^{-2}$ per pixel.

The Mayya et al. (1994) \ha data from which our fluxes were flux calibrated were not corrected for Galactic extinction.
Therefore, the \hii\ region luminosities of our catalogued \hii\ regions were corrected {\em a posteriori} for foreground 
Galactic extinction (Schlegel et al. 1998)\nocite{Schlegel1998} implying A$_{H\alpha}=$0.08~mag.

The bandwidth of the H$\alpha$ filter also contains emission from
the [\nii]$\lambda$6548\AA\ and [\nii]$\lambda$ 6584\AA\
emission lines at their corresponding red--shifted wavelengths. A proper
correction for this requires spectroscopic information for all the
\hii\ regions. This information is not currently available. 
The typical [\nii]/H$\alpha$ ratio for extragalactic \hii\ regions
of solar metallicity is approximately 0.33 (e.g. Kennicutt \& Kent 1983). Therefore, an approximate
correction of 25\% should be applied to the H$\alpha$ fluxes and
equivalent widths. 
Rather than applying this average correction to the fluxes of all \hii, we 
have opted to give the uncorrected H$\alpha$+[\nii]
luminosities in Table~2.

\subsection{\spitzer Infrared images}
To investigate the extinction and the star formation in the bar of NGC~2903 we make use of the \spitzer 24 \micron\ observations from the Multiband Imaging Photometer (MIPS) instrument  (Werner et al. 2004; Fazio et al. 2004; Rieke et al. 2004). The data was obtained in April 2004 as part of the Local Volume Legacy Survey (Lee at al. 2008; Kennicutt et al. 2007). MIPS reduction steps are described in detail by Gordon et al. (2005)\nocite{Gordon2005}. The field of view covered by the observations is large enough to contain the whole galaxy with enough sky coverage allowing a good sky subtraction.  The 24~\micron\  point--spread function (PSF) have a measured FWHM of $\sim$5.6'' and a $2\sigma$ sensitivity limit of $0.52 \times 10^{-6}$ Jy arcsec$^{-2}$.
The image was registered to the same coordinate system and pixel scale as the \ha image and the background was subtracted in the same way as for the \ha image. \nocite{werner,fazio,rieke,lee}

The 24~\micron\ image has a much larger PSF FWHM than the \ha image (FWHM  $\sim$2.1''). Therefore, to allow for accurate comparisons, we made a copy of the \ha image degraded to the 24 \micron\ PSF by convolving with a PSF kernel as described by Gordon et al. (2008)\nocite{Gordon2008}. This copy was only used to calculate the \ha attenuation of the \hii\ regions (Sec.~\ref{sec:dust}).
As complementary data we also use the 3.6 and 8\micron\ images, with PSF FWHM $\sim$2'' in both bands. These images are only used  to carry out global morphological comparisons with the rest of the  bands; therefore, we registered the images to the same coordinate system as the  \ha image, but we did not degraded their spatial resolution.

\subsection{\galex images}
To analyse the UV emission and its distribution in NGC~2903, we  use  \galex mission data (Martin et al. 2005\nocite{Martin2005}). The \galex direct imaging  observations are centered at 1529\AA{}  and at 2312\AA{} for the far-ultraviolet (FUV, 1350 - 1750\AA{}) and near-ultraviolet (NUV, 1750-2750\AA{}) bands respectively.
NGC~2903 was observed in February 2004 as a part of the {\em Nearby Galaxies Survey} (NGS, Bianchi et al. 2003\nocite{Bianchi2003}) with an exposure time of 1862s and 1861s for the FUV and NUV respectively. The PSF FWHM of the images are $\sim$4.2 and $\sim$4.6'' for the FUV and NUV respectively, and their respective  sensitivity limits $5.25\times10^{-18}$ and $3.09\times10^{-18}$~erg s$^{-1}$cm$^{-2}$\AA$^{-1}$ per pixel (see  Bianchi et al. 2003\nocite{Bianchi2003} for further details on the NGC~2903 \galex data).

Both the FUV and NUV images were aligned to the same coordinate system and pixel scale as the \ha image and the background was subtracted in the same way as 
for the \ha and \spitzer data.

\subsection{Other data: Ground--based optical images and CO~($J$=1-0)}
Ground based optical images from the {\em Sloan Digital Sky Survey} (SDSS) and a map of the CO~($J$=1-0) rotational transition emission have also been used for the morphological analysis of NGC~2903.
The optical data set comprises $r'$, $g'$ and $z'$ band images, which were  observed as part of the SDSS data release 6 and have a PSF FWHM of $\sim$ 1.1''.
We subtracted the sky emission from these images using the same IRAF script  used for the \ha image (Sect.~\ref{ha}).

\begin{figure*}
\centering 
\subfigure[NUV with overlaid \ha  countours.]{\includegraphics[scale= 0.3]{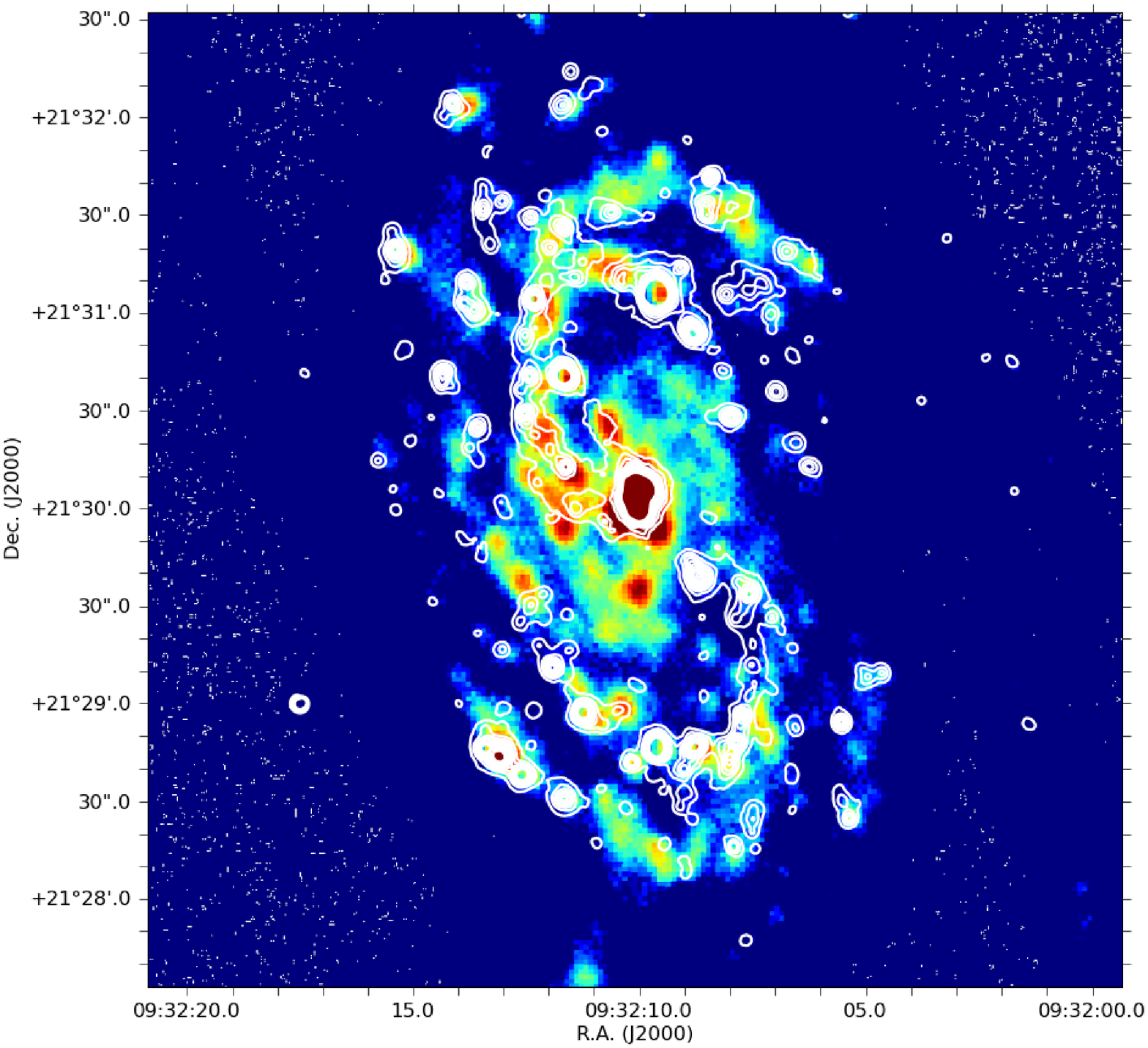}}
\subfigure[FUV with overlaid \ha contours.]{\includegraphics[scale = 0.3]{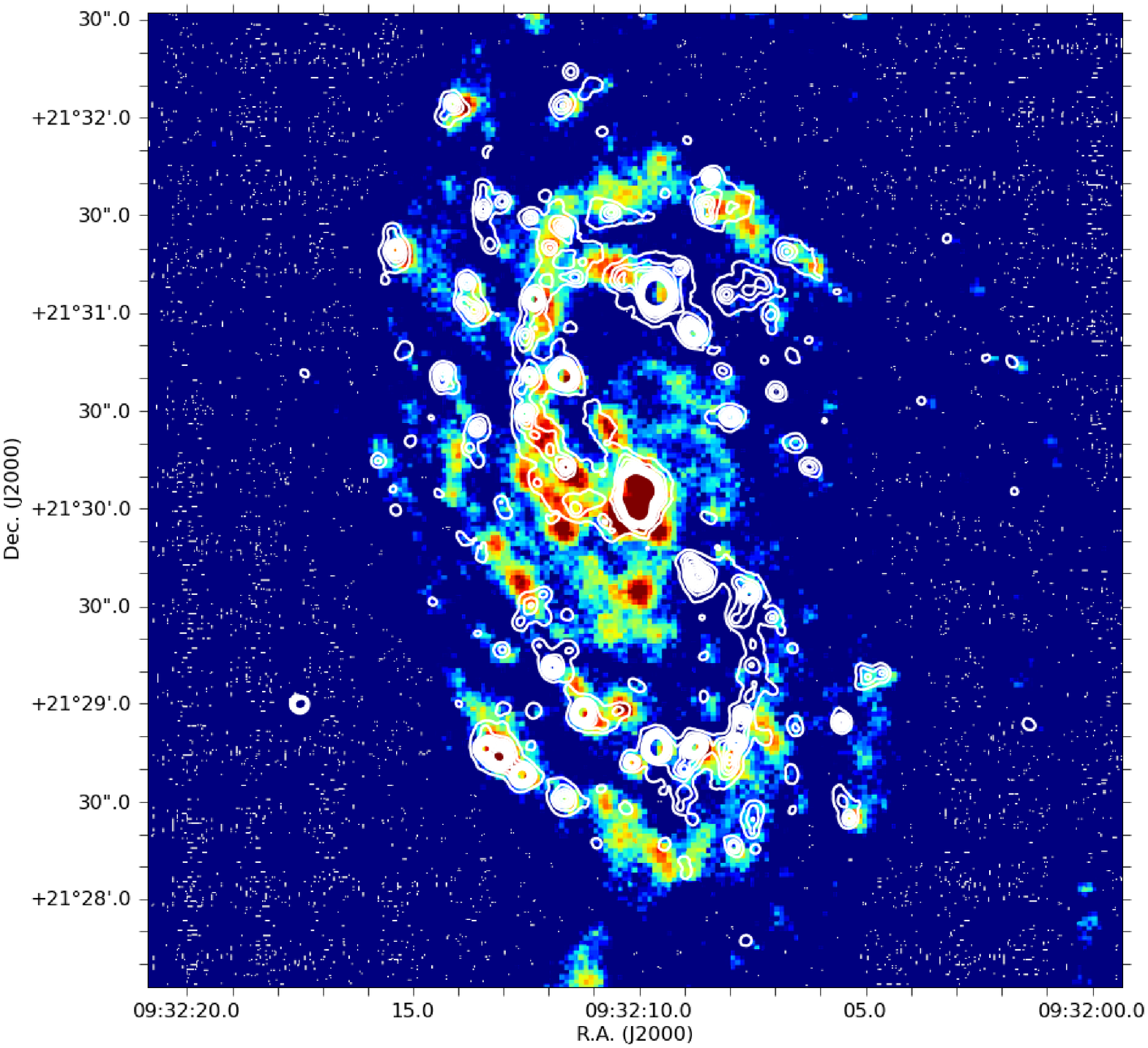}}
\subfigure[3.6 \micron\ with overlaid \ha contours.]{\includegraphics[scale = 0.3]{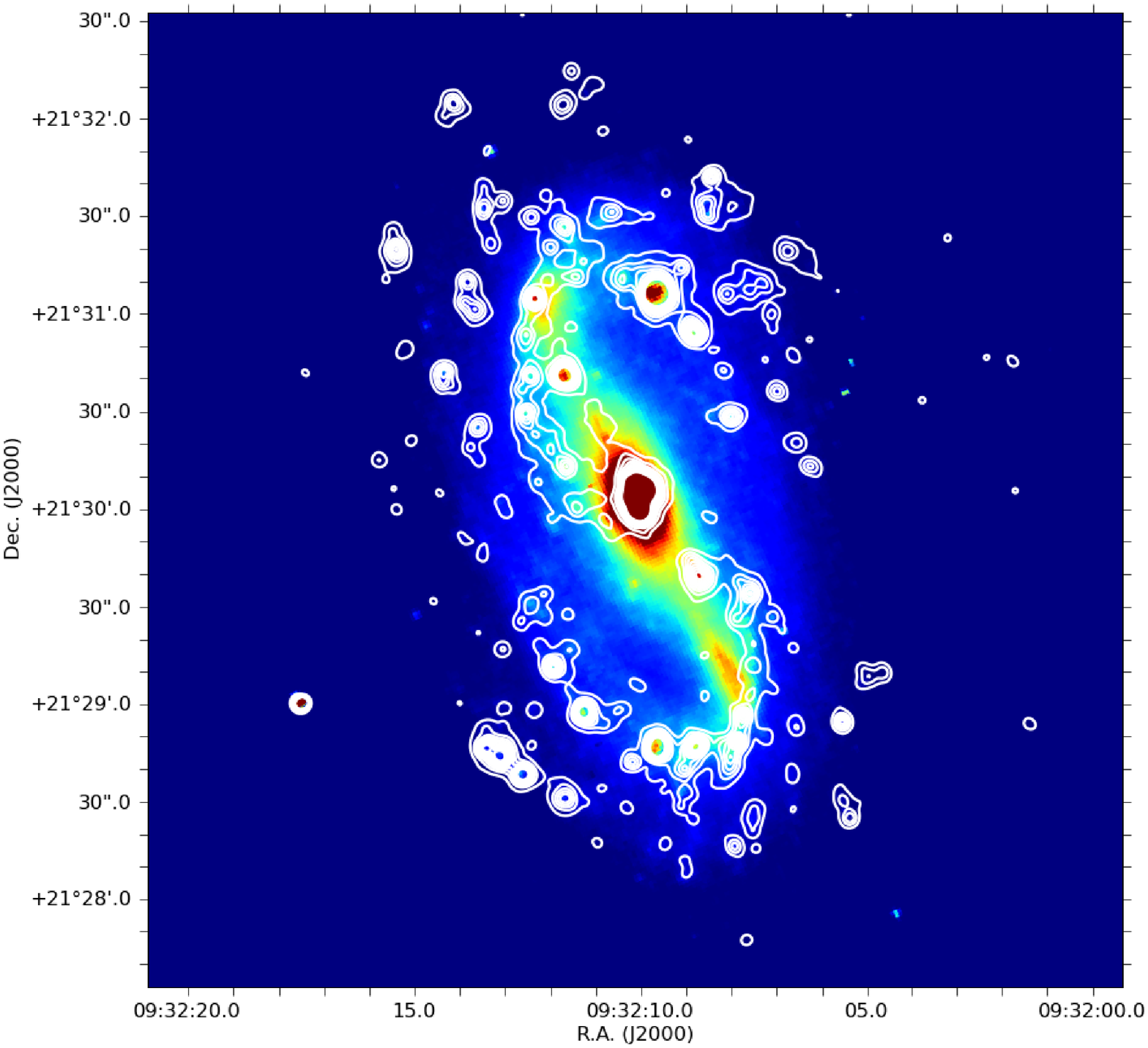}}
\subfigure[24 \micron\ with overlaid \ha contours.]{\includegraphics[scale = 0.3]{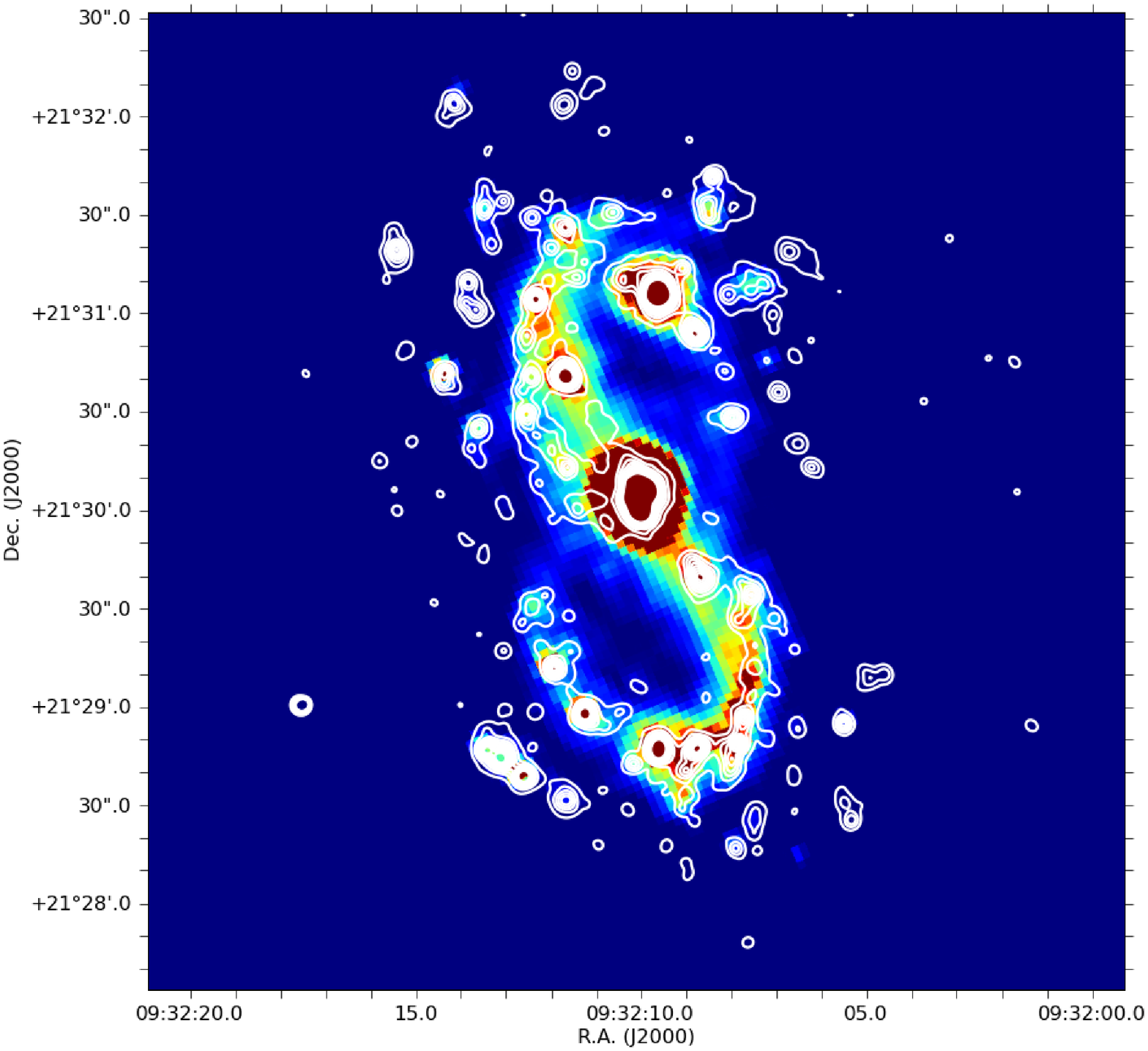}}
\subfigure[{\em g}'-{\em z}' colour map with overlaid \ha contours.]{\includegraphics[scale = 0.3]{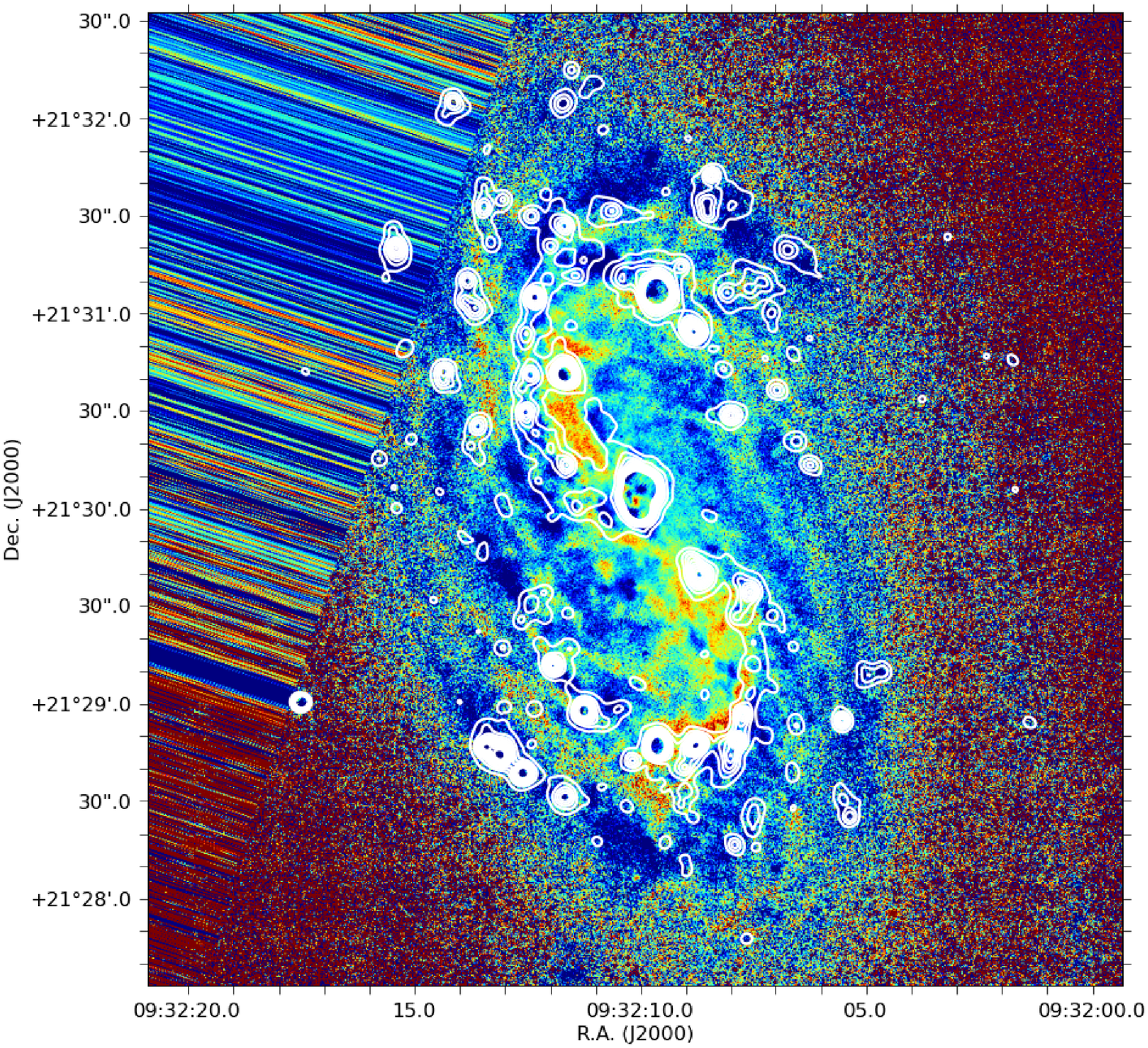}}
\vspace{-0.2cm}
\subfigure[CO~($J$=1-0) with overlaid  \ha contours.]{\includegraphics[scale = 0.3]{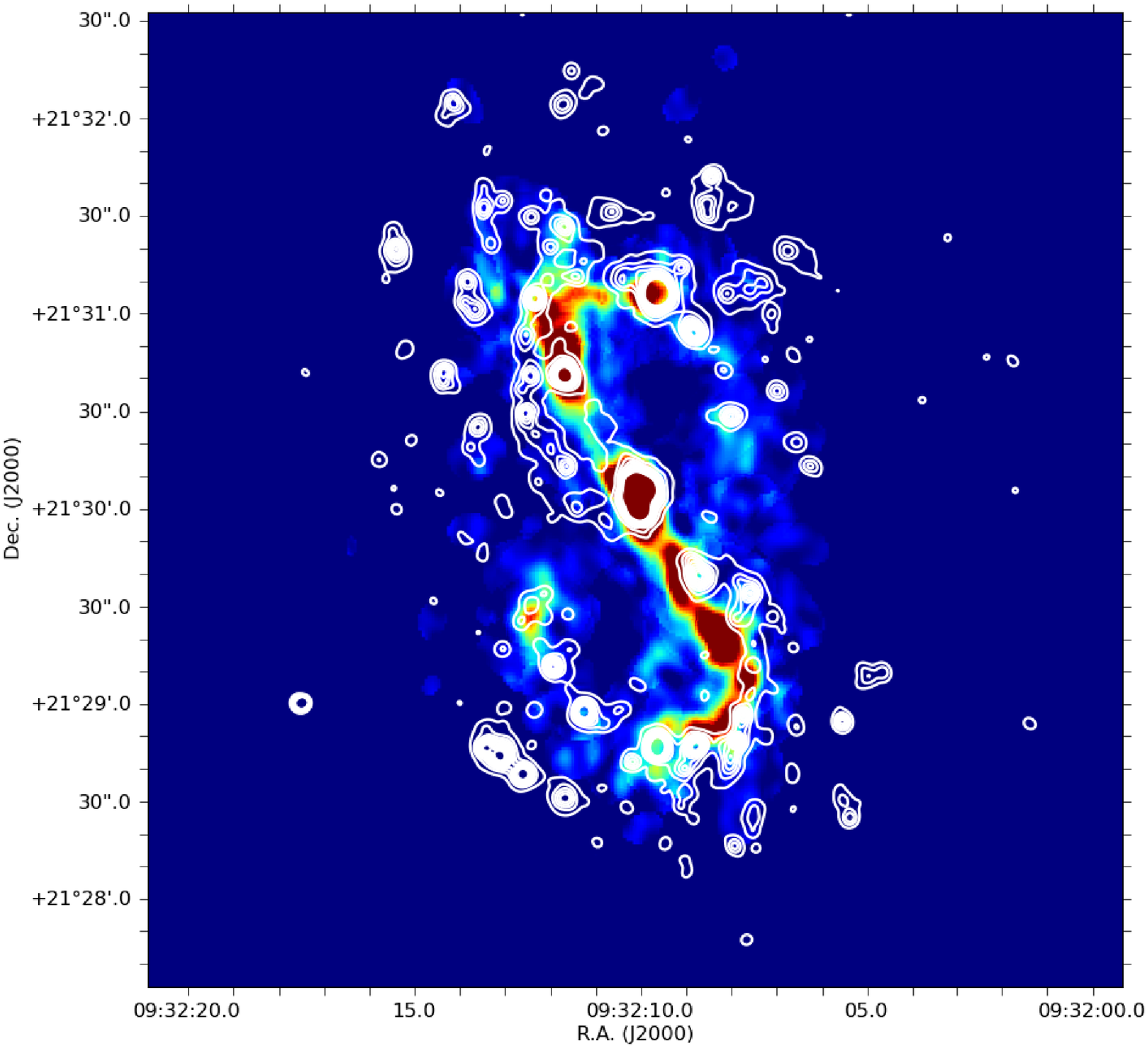}}
\caption{Images of NGC~2903 in different spectral bands: \textbf{(a)} \galex NUV image, logarithmic scale. \textbf{(b)} \galex FUV image, logarithmic scale,  \textbf{(c)} \spitzer 3.6$\mu$m, \textbf{(d)} \spitzer 24$\mu$m, \textbf{(e)} $g'-z'$ colour--map. \textbf{(f)} CO~($J$=1-0) transition map of NGC~2903. All the colour scales are given in arbitrary units. All images (a) to (f) show overlaid \ha contours, the resolution of each image is given in the text, Sect~\ref{ObservationsDataReduction}. The \ha contour levels range between 4.5$\times10^{35}$~erg/s/pix and 4.5$\times10^{36}$~erg/s/pix (the \ha\ image pixel size is 0.33''/pix). \label{fig:composition}}
\end{figure*}

For the photometric calibration, we estimated the calibration factor for converting the image digital counts to calibrated $g'$, $r'$ and $z'$ 
AB magnitudes as described on the  SDSS web page for the data release 6\footnote{http://www.sdss.org/dr6/algorithms/fluxcal.html}. 
These calibrated images were used to produce a $g'-z'$ colour map of NGC~2903 (Fig.~\ref{fig:composition}e). 

The CO~($J$=1-0) emission line map of NGC~2903 has been used to trace the molecular gas emission and was obtained from  the BIMA (Berkeley-Illinois-Maryl
and Association) array and the NRAO (National Radio Astronomy Observatory)\footnote{The National Radio Astronomy Observatory is a facility of the 
NSF operated under cooperative agreement by Associated Universities, Inc.} 12m single--dish telescope, as a part of the BIMA SONG key project. The emission line map has a maximum FWHM of 6.8''.
Data acquisition and reduction details can be found in Regan et al. 2001 and Helfer et al 2003.\nocite{Regan2001}  \nocite{Helfer2003}

%
%
%
%

\section{NGC 2903 general morphology}
\label{morphology}
The different appearance of galaxies in different wavelength ranges has already been illustrated by many authors. Galaxies in the UV bands seem to be patchier and later type than the same galaxies observed at optical and near infrared wavelengths. The stellar components such as bulges, bars and old stellar disks tend to disappear when observed in the UV range (e.g Kuchinski et al. 2000)\nocite{kuchinski}. NGC~2903 is a good example of this behaviour. Fig.~\ref{fig:composition} shows  NGC~2903 observed in different wavelength regimes:  FUV, NUV, 3.6~\micron\,  24~$\mu$m,  $g'-z'$ colour map and a CO~($J$=1-0) rotational transition emission line map. All images show  overlaid \ha contours.

The UV distribution is patchy and spiral-like and does not resemble the smooth bar-like distribution shown in the 3.6~\micron\ image. The CO~($J$=1-0) traces the 3.6~\micron\ emission, while the \ha follows an s--shape distribution also followed by the main dust--lanes, as shown by the $g'-z'$ colour image. Dust spurs emerge from the main dust lanes at many locations.

The \ha leads the CO~(J$=$1-0)  emission, taking into account that NGC~2903 rotates counter--clockwise (Hernandez et al. 2005), assuming trailing spiral arms.


\section{Photometry of star--forming regions}
\label{photometry}
\subsection{HII region catalogue}
\label{Ha_photometry}
\begin{figure}
 \includegraphics[scale = 0.43]{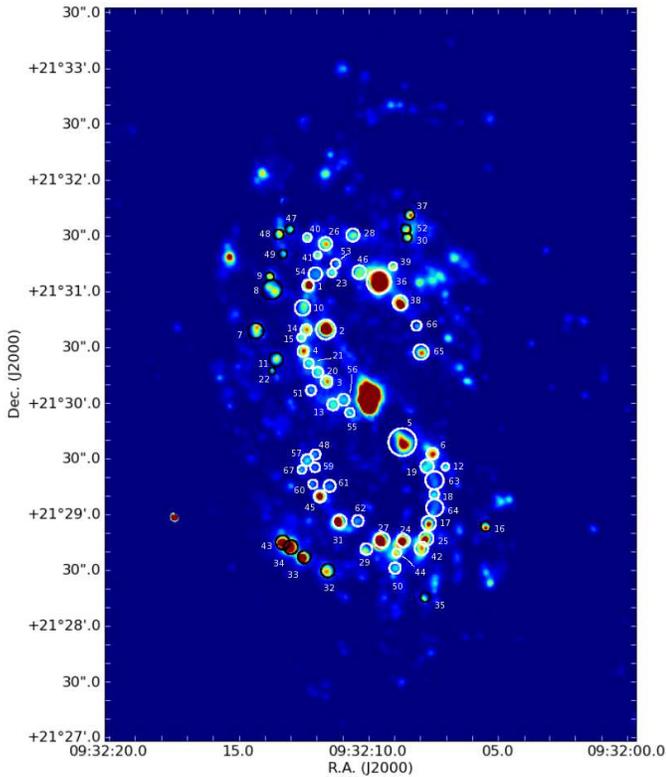}
\caption{\ha continuum--subtracted image of NGC~2903. The white open circles indicate the position and size of the integration apertures of the catalogued \hii\ regions.\label{fig:HII_catalogue} belonging to the bar region. The black open circles indicate the \hii\ regions closest to the catalogued UV complexes without \ha counterparts.}
\end{figure}

To study the properties of the recent star forming sites in the NGC~2903
bar we produced a catalogue  of the \hii\ regions located in the bar
zone (except the nucleus). We consider the bar zone as the area covering the whole {\em lense region} (Kormendy 1979\nocite{Kormendy79}.) as observed in the 3.6~\micron\ image, 
see Fig.~\ref{fig:composition}c.
The selection criteria for considering a feature in the \ha image as an
\hii\ region is that the feature must have an area in pixels equal or
larger than the image spatial resolution (i.e. area$\approx$32 pixels; with a pixel size of 0.33''/pix),
having all pixels an intensity of at least three times the r.m.s. noise
above the local background intensity level. This selection criteria implies a
detectability limit of $\sim 8.3 \times 10^{37}$  erg~s$^{-1}$. \\
A total of 67 \hii\ regions were catalogued and their positions, sizes
and luminosities were measured. We performed aperture photometry using
GAIA\footnote{GAIA is a derivative of the Skycat catalogue and image 
display tool, developed as part of the VLT project at ESO. Skycat and 
GAIA are free software under the terms of the GNU copyright.}.
Most of the catalogued regions are located in the s-shape region and its
neighbourhood, a region which is closely related to the stellar bar.
A representation of the catalogued \hii\ regions can be seen in Fig.~\ref{fig:HII_catalogue}.

The \hii\ region catalogue comprises the position, size, shape and H$\alpha$+[\nii]
luminosity of the regions, and it is presented in Table~\ref{table:HII}.
The table contains the \hii\ region identification number in column 1;
the equatorial coordinates (columns 2 and 3). Column 4 shows the radius
in arcseconds; columns 5 and 6 the eccentricity and position angle
(measured North-West) of the integration aperture, and column 7
shows the decimal logarithm of the Galactic extinction corrected H$\alpha$+[\nii] luminosity 
(not corrected for internal dust extinction). 

\subsubsection{\ha equivalent width}
The \ha EW (hereinafter EW$_{H\alpha}$) is a measure of the strength of the 
\ha emission with respect to the continuum emission of the stellar 
ionizing cluster. Therefore, the EW$_{H\alpha}$ depends mainly on the evolutionary
stage of the \hii\ region (decreasing as the region ages), 
but also on the metallicity, the IMF, dust content and 
ionizing photon leakage (e.g. Bresolin \& Kennicutt 1999; Zurita \& P\'erez 2008).

The \ha EWs were calculated from the ratio of our measured 
\ha luminosities (not corrected for internal dust extinction) to the 
continuum emission per \AA. The continuum emission of the ionizing 
cluster has been estimated from the broad band emission in 
the R--band at the position of the catalogued \hii\ regions.
Rather than using our  Johnson R--band image of NGC~2903, we 
used the Sloan Digital Sky Survey (SDSS) $r'$--band image. The latter has
a higher spatial resolution than our image, and therefore allowed us to
better identify and measure the continuum cluster emission.

When computing the \ha EW of an \hii\ region we are interested only in
the continuum radiation emitted by the ionizing cluster. However, both
the flux  from the ionizing cluster and the continuum emission from the
underlying  stellar population of the galaxy are included within the
integration  apertures. Correcting for this contamination is not
straightforward and it is the major source of uncertainty in \hii\
region broad--band fluxes (Zurita \& P\'erez 2008).
We followed the procedure described in Zurita \& P\'erez (2008) 
to estimate the contribution of the underlying stellar 
population to our measured broad band fluxes by using two of 
the methods described there.
The first method consist of the measurement of the local background 
from the median value within annular apertures around the \hii\ regions.
The inner radius was set to the effective radius of the \hii\ region,
and the annulus width to approximately three times the  \hii\ region
radius. 
The areas of the image corresponding to catalogued \hii\ regions had 
previously been masked out in the image to avoid contamination 
within the annular apertures from neighbouring \hii\ regions.
For some regions, the median continuum emission within  the annulus 
is clearly overestimated and yields an underestimation of the cluster
continuum emission, which sometimes even results in negative
fluxes. This is generally due to strong spatial variation of the disk
continuum emission at small scales. Therefore, a second approach was adopted for better constraining the
local background estimates.

The second method uses growth curves and takes advantage of the fact that
in most \hii\ regions the continuum emission is less extended 
than the H$\alpha$ emitting area. Therefore, the local
background can be obtained from within the \hii\ region area as defined
by its H$\alpha$ emission. For each \hii\ region we derived 
radially averaged $r'$--band surface--brightness  profiles. For 
each region we selected a radial range not contaminated 
by the ionizing cluster and determined the local background surface
brightness from a fit to the profile in that range.

The $r'$--band continuum fluxes were all corrected for the underlying stellar population contamination using both methods (when possible), 
giving   us two estimates for the $r'$--band flux of the ionizing cluster, from which we derived two values for the EW$_{H\alpha+[NII]}$.
We calculated the difference between the two EW$_{H\alpha+[NII]}$ estimates for each \hii\ region, and this was taken as the measured 
H$\alpha$+[\nii] EW uncertainty, so it represents the range of values covered when using the different estimates.
Those regions for which the two estimates yielded differences larger than 600\AA,  and those regions for which only one background measurement was available, were discarded, as we considered the results non-reliable.
The resulting \ha\ EWs (including emission from the \nii\ lines) are presented in column 8 of Table~\ref{table:HII}.
\begin{table*}[!]
\vspace*{-0.30cm}
\caption{\tiny \hii\ region catalogue of the bar zone of NGC~2903. Column 1: ID number of the catalogued \hii\ regions; Cols. 2 and 3: equatorial coordinates; Col. 4: \hii\ region radius in arcseconds; Cols. 5 and 6: Eccentricity and position angle (measured North-West) of the integration apertures; Col. 7: decimal logarithm of the \ha luminosity (in erg~s$^{-1}$); Col. 8:  decimal logarithm of the \ha equivalent width; Col. 9: decimal logarithm of the 24~\micron\  emission within the \hii\ region apertures; Col. 10 \ha attenuation  see Sect.~\ref{sec:dust}; Col. 11: estimate of the current star formation rate for each catalogued \hii\ region. Cols. 7,8,9 and 10 include the corresponding correction for Galactic extinction (Schlegel et al. 1998)\nocite{Schlegel1998}. The SFRs have been obtained after applying a correction of 25\% for [\nii] contamination to the \ha luminosities and dust extinction as given in column 10.\label{table:HII}}
\vspace*{-0.35cm}
\begin{center}
\begin{tabular}{ccccccccccc}
\hline
ID   &    R.A.  & Decl.  & Radius   & $e$    & P.A.     & $\log$~L$_{H_{\alpha}+[NII]}$ & $\log$~EW$_{H_{\alpha}+[NII]}$ &   $\log(L_{24\mu m})$ & A$_{H\alpha}$ & SFR  \\
     & (J2000) & (J2000) & (arcsec) &        &  ($\deg$)& (in erg~s$^{-1}$)      &    (in \AA)          &   (erg~s$^{-1}$) & (mag) & ($10^{-3} M_\odot yr^{-1}$)\\	
\hline
  1    &  09:32:12.35  &  21:31:04.66  &  3.1  &  0  &  90  &    38.99 $\pm$ 0.05    &    2.32 $\pm$ 0.11    &  40.60  &  1.21  &    17.7 $\pm$ 0.9 \\
2    &  09:32:11.69  &  21:30:40.84  &  5.23  &  0.51  &  158  &    39.41 $\pm$ 0.05    &    2.86 $\pm$ 0.02    &  40.93  &  1.07  &    41.0 $\pm$ 2.2 \\
3    &  09:32:11.67  &  21:30:12.75  &  3.17  &  0  &  0  &    38.71 $\pm$ 0.10    &   -   &  40.24  &  1.09  &    8.3 $\pm$ 0.9 \\
4    &  09:32:12.56  &  21:30:29.23  &  3  &  0  &  0  &    38.75 $\pm$ 0.08    &    2.70 $\pm$ 0.09    &  40.19  &  0.95  &    7.9 $\pm$ 0.8 \\
5    &  09:32:08.77  &  21:29:39.86  &  7.3  &  0.81  &  -28  &    39.28 $\pm$ 0.08    &    2.34 $\pm$ 0.04    &  40.71  &  0.93  &    26.5 $\pm$ 2.9 \\
6    &  09:32:07.62  &  21:29:33.54  &  3.1  &  0  &  0  &    38.81 $\pm$ 0.08    &    3.07 $\pm$ 0.03    &  40.35  &  1.09  &    10.5 $\pm$ 0.9 \\
7    &  09:32:14.38  &  21:30:40.38  &  4.23  &  0.68  &  6  &    38.76 $\pm$ 0.12    &   -   &  40.24  &  1.01  &    8.6 $\pm$ 1.2 \\
8    &  09:32:13.75  &  21:31:02.56  &  5.23  &  0.8  &  -48  &    38.87 $\pm$ 0.11    &   -   &  39.49  &  0.21  &    5.3 $\pm$ 1.5 \\
9    &  09:32:13.83  &  21:31:09.62  &  2.33  &  0  &  0  &    38.43 $\pm$ 0.10    &    2.50 $\pm$ 0.01    &  39.32  &  0.36  &    2.2 $\pm$ 0.5 \\
10   &  09:32:12.57  &  21:30:52.67  &  4.17  &  0.77  &  10  &    38.60 $\pm$ 0.14    &   -   &  40.05  &  0.96  &    5.7 $\pm$ 1.0 \\
11   &  09:32:13.60  &  21:30:24.86  &  3.7  &  0.52  &  38  &    38.62 $\pm$ 0.14    &   -   &  40.01  &  0.88  &    5.6 $\pm$ 1.1 \\
12   &  09:32:07.12  &  21:29:26.37  &  2.03  &  0  &  0  &    38.12 $\pm$ 0.16    &    2.28 $\pm$ 0.01    &  39.65  &  1.08  &    2.1 $\pm$ 0.4 \\
13   &  09:32:11.43  &  21:30:00.36  &  3.1  &  0  &  0  &    38.55 $\pm$ 0.14    &   -   &  39.97  &  0.92  &    4.9 $\pm$ 0.9 \\
14   &  09:32:12.44  &  21:30:40.75  &  3  &  0  &  0  &    38.70 $\pm$ 0.09    &    2.43 $\pm$ 0.14    &  40.13  &  0.93  &    7.0 $\pm$ 0.8 \\
15   &  09:32:12.63  &  21:30:36.70  &  2.23  &  0  &  0  &    38.30 $\pm$ 0.13    &   -   &  39.70  &  0.90  &    2.7 $\pm$ 0.5 \\
16   &  09:32:05.59  &  21:28:54.01  &  2.8  &  0  &  90  &    38.59 $\pm$ 0.10    &    2.87 $\pm$ 0.03    &  39.77  &  0.61  &    4.1 $\pm$ 0.7 \\
17   &  09:32:07.77  &  21:28:56.02  &  3.93  &  0  &  0  &    38.87 $\pm$ 0.11    &   -   &  40.51  &  1.27  &    14.1 $\pm$ 1.4 \\
18   &  09:32:07.56  &  21:29:11.66  &  2.6  &  0  &  0  &    38.33 $\pm$ 0.16    &   -   &  40.23  &  1.74  &    6.4 $\pm$ 0.6 \\
19   &  09:32:07.83  &  21:29:26.69  &  3.57  &  0  &  90  &    38.66 $\pm$ 0.14    &   -   &  40.32  &  1.30  &    9.0 $\pm$ 1.2 \\
20   &  09:32:12.01  &  21:30:17.87  &  2.87  &  0.85  &  3  &    38.22 $\pm$ 0.13    &   -   &  39.79  &  1.14  &    2.8 $\pm$ 0.4 \\
21   &  09:32:12.35  &  21:30:22.63  &  2.7  &  0.78  &  108  &    38.25 $\pm$ 0.13    &    2.60 $\pm$ 0.09    &  39.49  &  0.68  &    2.0 $\pm$ 0.4 \\
22   &  09:32:13.76  &  21:30:18.72  &  1.9  &  0  &  0  &    38.00 $\pm$ 0.18    &   -   &  39.22  &  0.66  &    1.1 $\pm$ 0.3 \\
23   &  09:32:11.44  &  21:31:11.46  &  2.33  &  0  &  90  &    38.30 $\pm$ 0.14    &   -   &  39.87  &  1.15  &    3.4 $\pm$ 0.5 \\
24   &  09:32:08.76  &  21:28:46.76  &  3.53  &  0  &  90  &    39.10 $\pm$ 0.05    &    2.31 $\pm$ 0.04    &  40.50  &  0.89  &    16.8 $\pm$ 1.2 \\
25   &  09:32:07.88  &  21:28:47.69  &  3.93  &  0  &  0  &    39.02 $\pm$ 0.07    &    2.63 $\pm$ 0.01    &  40.59  &  1.15  &    18.0 $\pm$ 1.4 \\
26   &  09:32:11.67  &  21:31:27.14  &  3.6  &  0  &  0  &    38.77 $\pm$ 0.11    &   -   &  40.39  &  1.23  &    10.8 $\pm$ 1.2 \\
27   &  09:32:09.59  &  21:28:47.12  &  4.17  &  0  &  0  &    39.41 $\pm$ 0.03    &    2.44 $\pm$ 0.02    &  40.93  &  1.08  &    40.8 $\pm$ 1.6 \\
28   &  09:32:10.62  &  21:31:31.74  &  3.5  &  0  &  90  &    38.62 $\pm$ 0.15    &    2.56 $\pm$ 0.04    &  40.20  &  1.17  &    7.2 $\pm$ 1.1 \\
29   &  09:32:10.18  &  21:28:42.26  &  3.1  &  0  &  90  &    38.52 $\pm$ 0.15    &    2.31 $\pm$ 0.03    &  40.13  &  1.21  &    6.0 $\pm$ 0.9 \\
30   &  09:32:08.53  &  21:31:30.16  &  2.53  &  0  &  0  &    38.46 $\pm$ 0.11    &    2.44 $\pm$ 0.06    &  40.10  &  1.27  &    5.5 $\pm$ 0.6 \\
31   &  09:32:11.20  &  21:28:57.28  &  3.67  &  0  &  0  &    39.19 $\pm$ 0.04    &    2.67 $\pm$ 0.03    &  40.59  &  0.89  &    20.7 $\pm$ 1.2 \\
32   &  09:32:11.67  &  21:28:30.75  &  3.7  &  0  &  0  &    38.84 $\pm$ 0.10    &    2.67 $\pm$ 0.07    &  39.62  &  0.29  &    5.3 $\pm$ 1.3 \\
33   &  09:32:12.57  &  21:28:38.22  &  3.5  &  0  &  0  &    39.14 $\pm$ 0.04    &    2.76 $\pm$ 0.02    &  40.56  &  0.91  &    19.1 $\pm$ 1.1 \\
34   &  09:32:13.09  &  21:28:43.61  &  3.83  &  0  &  0  &    39.24 $\pm$ 0.04    &    2.49 $\pm$ 0.03    &  40.19  &  0.41  &    14.8 $\pm$ 1.4 \\
35   &  09:32:07.93  &  21:28:16.18  &  2.67  &  0  &  0  &    38.34 $\pm$ 0.17    &    2.63 $\pm$ 0.00    &  39.73  &  0.88  &    2.9 $\pm$ 0.7 \\
36   &  09:32:09.63  &  21:31:06.38  &  6.4  &  0  &  90  &    39.81 $\pm$ 0.03    &    2.51 $\pm$ 0.02    &  41.47  &  1.29  &    126.1 $\pm$ 3.8 \\
37   &  09:32:08.43  &  21:31:42.49  &  2.8  &  0  &  90  &    38.61 $\pm$ 0.10    &   -   &  39.67  &  0.50  &    3.8 $\pm$ 0.7 \\
38   &  09:32:08.81  &  21:30:55.01  &  4  &  0  &  0  &    39.14 $\pm$ 0.06    &    2.52 $\pm$ 0.08    &  40.51  &  0.85  &    17.8 $\pm$ 1.5 \\
39   &  09:32:09.08  &  21:31:14.75  &  2.17  &  0  &  0  &    38.35 $\pm$ 0.11    &   -   &  40.07  &  1.41  &    4.9 $\pm$ 0.4 \\
40   &  09:32:12.38  &  21:31:30.57  &  2.33  &  0  &  0  &    38.30 $\pm$ 0.14    &   -   &  38.98  &  0.24  &    1.5 $\pm$ 0.5 \\
41   &  09:32:11.97  &  21:31:21.00  &  2  &  0  &  0  &    38.22 $\pm$ 0.12    &   -   &  39.68  &  0.98  &    2.4 $\pm$ 0.4 \\
42   &  09:32:08.06  &  21:28:42.43  &  3.67  &  0  &  0  &    38.87 $\pm$ 0.09    &    2.25 $\pm$ 0.01    &  40.32  &  0.97  &    10.7 $\pm$ 1.2 \\
43   &  09:32:13.40  &  21:28:46.21  &  3.83  &  0  &  0  &    39.16 $\pm$ 0.05    &    2.41 $\pm$ 0.03    &  40.20  &  0.48  &    13.4 $\pm$ 1.4 \\
44   &  09:32:09.01  &  21:28:40.08  &  2.37  &  0  &  0  &    38.52 $\pm$ 0.09    &   -   &  40.38  &  1.67  &    9.2 $\pm$ 0.5 \\
45   &  09:32:11.96  &  21:29:10.88  &  3.27  &  0  &  0  &    38.90 $\pm$ 0.07    &    2.64 $\pm$ 0.02    &  40.55  &  1.27  &    15.3 $\pm$ 1.0 \\
46   &  09:32:10.39  &  21:31:11.67  &  3.7  &  0  &  0  &    38.79 $\pm$ 0.11    &    2.43 $\pm$ 0.04    &  40.42  &  1.25  &    11.5 $\pm$ 1.3 \\
47   &  09:32:13.05  &  21:31:35.10  &  2.43  &  0  &  0  &    38.28 $\pm$ 0.16    &    2.85 $\pm$ 0.01    &  39.45  &  0.62  &    2.0 $\pm$ 0.5 \\
48   &  09:32:13.47  &  21:31:32.35  &  2.7  &  0  &  0  &    38.53 $\pm$ 0.11    &   -   &  40.00  &  0.99  &    5.0 $\pm$ 0.7 \\
49   &  09:32:13.32  &  21:31:21.74  &  2.4  &  0  &  0  &    38.19 $\pm$ 0.19    &    2.72 $\pm$ 0.09    &  39.47  &  0.73  &    1.8 $\pm$ 0.5 \\
50   &  09:32:09.09  &  21:28:31.92  &  3.03  &  0  &  0  &    38.42 $\pm$ 0.18    &   -   &  40.34  &  1.80  &    8.1 $\pm$ 0.9 \\
51   &  09:32:12.27  &  21:30:08.09  &  2.7  &  0.03  &  83  &    38.21 $\pm$ 0.23    &   -   &  39.72  &  1.04  &    2.5 $\pm$ 0.7 \\
52   &  09:32:08.58  &  21:31:34.64  &  2.67  &  0  &  0  &    38.56 $\pm$ 0.10    &   -   &  40.10  &  1.09  &    5.9 $\pm$ 0.7 \\
53  &  09:32:11.28  &  21:31:16.48  &  2.5  &  0  &  0  &   38.17 $\pm$ 0.21   &   2.21 $\pm$ 0.36   &  39.93  &  1.48  &   3.4 $\pm$ 0.6 \\
54  &  09:32:12.06  &  21:31:11.01  &  3.67  &  0  &  0  &   38.53 $\pm$ 0.20   &   1.99 $\pm$ 0.11   &  40.29  &  1.47  &   7.8 $\pm$ 1.2 \\
55  &  09:32:10.77  &  21:29:56.08  &  2.5  &  0  &  0  &   38.27 $\pm$ 0.17   &   1.85 $\pm$ 0.21   &  40.29  &  2.00  &   6.9 $\pm$ 0.6 \\
56  &  09:32:11.01  &  21:30:03.02  &  3.2  &  0  &  0  &   38.51 $\pm$ 0.16   &   -   &  40.63  &  2.21  &   14.6 $\pm$ 0.9 \\
57  &  09:32:12.42  &  21:29:30.64  &  3.17  &  0  &  0  &   38.47 $\pm$ 0.17   &   -   &  40.11  &  1.26  &   5.6 $\pm$ 0.9 \\
58  &  09:32:12.10  &  21:29:33.70  &  2.67  &  0  &  0  &   38.18 $\pm$ 0.24   &   3.15 $\pm$ 0.07   &  39.70  &  1.06  &   2.4 $\pm$ 0.7 \\
59  &  09:32:12.12  &  21:29:26.71  &  2.6  &  0  &  0  &   38.12 $\pm$ 0.26   &   -   &  39.78  &  1.30  &   2.6 $\pm$ 0.6 \\
60  &  09:32:12.21  &  21:29:17.71  &  2.43  &  0  &  0  &   38.06 $\pm$ 0.26   &   -   &  39.92  &  1.67  &   3.2 $\pm$ 0.5 \\
61  &  09:32:11.56  &  21:29:16.67  &  3.2  &  0  &  0  &   38.32 $\pm$ 0.25   &   2.85 $\pm$ 0.06   &  40.03  &  1.38  &   4.4 $\pm$ 0.9 \\
62  &  09:32:10.49  &  21:28:57.76  &  3.07  &  0.55  &  61  &   38.27 $\pm$ 0.21   &   2.29 $\pm$ 0.05   &  39.43  &  0.60  &   1.9 $\pm$ 0.7 \\
63  &  09:32:07.53  &  21:29:19.29  &  4.9  &  0.9  &  -18  &   38.33 $\pm$ 0.25   &   2.46 $\pm$ 0.04   &  40.10  &  1.50  &   5.0 $\pm$ 1.0 \\
64  &  09:32:07.53  &  21:29:04.49  &  4.63  &  0.87  &  14  &   38.37 $\pm$ 0.23   &   -   &  40.16  &  1.54  &   5.8 $\pm$ 1.0 \\
65  &  09:32:08.01  &  21:30:28.55  &  4.13  &  0  &  90  &   38.83 $\pm$ 0.13   &   -   &  39.96  &  0.56  &   6.7 $\pm$ 1.6 \\
66  &  09:32:08.18  &  21:30:42.83  &  2.67  &  0  &  0  &   38.16 $\pm$ 0.25   &   -   &  39.09  &  0.39  &   1.2 $\pm$ 0.7 \\
67  &  09:32:12.63  &  21:29:25.61  &  2.33  &  0  &  0  &   38.13 $\pm$ 0.21   &   -   &  39.50  &  0.86  &   1.7 $\pm$ 0.5 \\
\hline
\hline
\end{tabular}
\end{center}
\end{table*}


\subsection{UV  photometry}
\label{UV_catalogue}

\begin{figure}[!b]
\includegraphics[scale = 0.47]{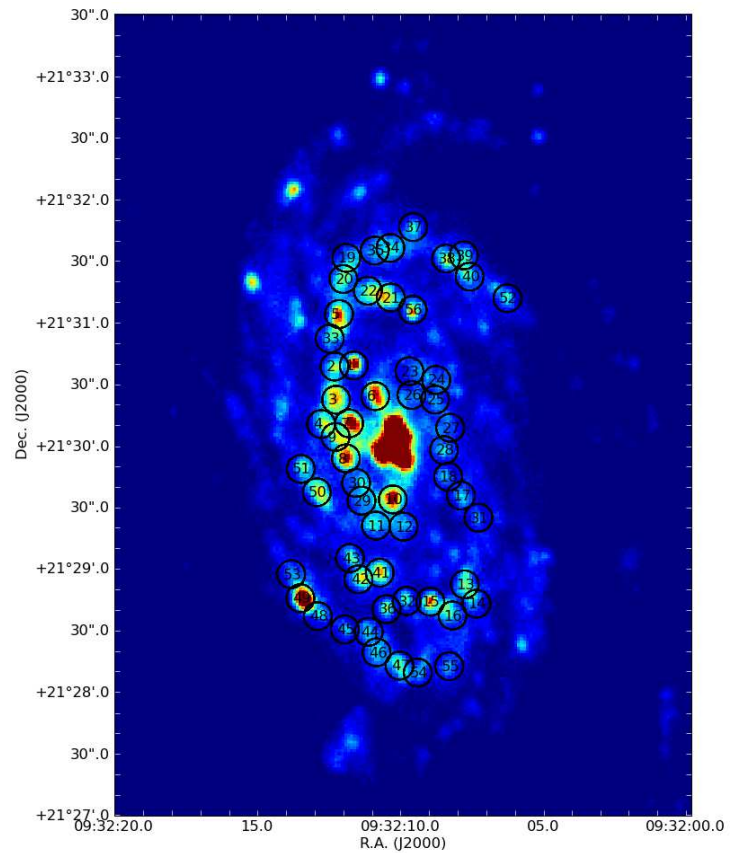}
\caption{\label{fig:regions}\galex NUV image of NGC~2903 with the apertures used for photometry in the FUV, NUV, \ha and 24~\micron\ images. See Sect.~\ref{UV_catalogue} for further details.}
\end{figure}

We performed aperture photometry on the NUV and FUV \galex images to catalogue all the UV emitting knots of the bar region with the aim of obtaining their luminosities and colours.
The aperture photometry has been carried out as described in Sect.~\ref{Ha_photometry} for the \hii\ region catalogue, but using circular apertures centered on the FUV peaks. The location of the apertures is shown in Fig.~\ref{fig:regions}. The same apertures and centers have been used to measure fluxes on the \ha and  24~\micron\  with the aim of deriving the dust attenuation (Sect.~\ref{sec:dust}). The aperture size selection, 13.2 arcsec, which corresponds to $\sim$570 pc at the galaxy distance, was determined by the spatial resolution of the 24~\micron\ image and the emission peak shifts of the emitting knots at the different bands (see below in this section).
These large apertures can cause some overlap between integration areas of neighbouring regions, but in none of the cases the contamination is significant.

As already mentioned for the \ha photometry in Sect.~\ref{Ha_photometry}, the background emission has to be subtracted from the measured 24~\micron, NUV and FUV  fluxes within the aperture, to yield the  stellar cluster emission alone.
For estimating the background contamination,  annuli around each aperture are difficult to define  because of crowding. As an alternative method, we fit light profiles perpendicular to the bar major axis averaged  over a rectangle of width $\approx$ 40'' (much larger than our photometric apertures for averaging out small scale background variations, but small enough to ensure a valid local background estimate). The local background of each region is then obtained by interactive interpolation of the background on the corresponding fitted profile (second order polynomial) at the position of the emission peaks.
A total of 56 regions were catalogued in the bar region alone (excluding the galaxy center) and the positions, luminosities and UV colours are shown in Table~\ref{table:attenuation}.

The peak emission in the UV and in the 24 \micron\ (and \ha) image is displaced a few arcseconds with respect to each other in most of the regions. This offset between the emission peaks in UV and 24 \micron\ or \ha amounts up to 6'' and was already noticed by Calzetti et al. (2005) for M51. 

As already said, the 24~\micron\ and \ha emission peaks do not always fall within the UV apertures. Therefore, the \ha\ and 
24~\micron\ luminosities shown in Table~\ref{table:attenuation} should be used with care, as they show the luminosities of regions centered on the UV peaks with the unique aim of obtaining the dust attenuation in the UV. For the \ha\ luminosities of the \hii\ regions we refer the reader to Sect.~\ref{Ha_photometry} and Table~\ref{table:HII}. Last column of Table~\ref{table:attenuation} shows the ID number of the \hii\ region catalogue, matching to the UV emitting knot.

A careful visual  inspection of the UV, the \ha and the 24~\micron\ images of NGC~2903, shows that a number of compact bright UV emission knots, mainly located in the bar region, in the Southeastern and Northwestern area from the nucleus, do not have significant \ha nor a 24~\micron\ counterpart emission (see Fig.~\ref{fig:composition}).

\subsection{Dust attenuation}
\label{sec:dust}

Since many years a lot of effort has been put into deriving reliable and easy to use dust extinction indicators to correct optical and UV observations (e.g. Kennicutt 1983; Calzetti et al. 2000; Iglesias-P\'aramo et al. 2006, etc).\nocite{kennicutt1983b,calzetti2000,iglesias2006} The appearance of  multi-wavelength observations covering from the FUV to the mid-infrared of a relatively large sample of nearby galaxies (e.g. SINGS; Kennicutt et al. 2003) \nocite{kennicutt2003} has allowed to us determine reliably extinction correction using all the available bands. The \ha and 24~\micron\ combination has shown to be a very useful tool to estimate attenuation corrected \ha and UV fluxes (Kennicutt et al. 2007; Kennicutt et al. 2009).\nocite{kennicutt2009} We have used these results to obtain dust--corrected UV fluxes. We calculate the attenuation in the UV bands from the H$\alpha$ attenuation using the laws presented by Calzetti (2001)\nocite{Calzetti2001}:
$$A_{0.16,star}=1.78A_{H\alpha,gas}\;\;\;(FUV)$$
$$A_{0.28,star}=1.29A_{H\alpha,gas}\;\;\;(NUV)$$
where $A_{X}$ is the attenuation, in magnitudes, at the specific band. To obtain the \ha attenuation we used the method described by Kennicutt et al. (2007) \nocite{Kennicutt2007}: 
$$A_{H\alpha} = 2.5\log \biggl[ 1 + \frac{aL_{24\mu m}}{(L_{H\alpha})_{obs}}\biggr]$$
where $(L_{H \alpha})_{obs}$ refers to the observed  H$\alpha$ luminosity with correction for the \nii\ lines, $L_{24\mu m}$ is defined as the product $\nu L_{\nu}$ and  $a$ is the $L_{24\mu m}/L_{H\alpha}$ constant scaling factor, which is fitted empirically. A value of 0.038 was adopted for $a$, as derived in the same paper for the derivation of individual region parameters, and a value of a = 0.02 was adopted when deriving galaxy integrated parameters. Galactic extinction was applied for the calculation of the dust attenuation.
Errors in the attenuation and UV magnitudes follow from two uncertainties; first, the photometric calibration uncertainties and second, the local background estimates. Errors are of the order of 10\%.

Because no dust tracers are detected for the UV emitting regions without  \ha and 24~\micron\ emission counterparts, we adopted a value of zero for the attenuation in these regions. Besides the lack of \ha and 24~\micron, no 8\micron\ (Dale et al. 2009)\nocite{dale2009} nor CO~($J$=1-0) emission is detected in these regions, indicating that there is very little gas or dust obscuring the UV emission. As an  $L_{H\alpha}$ upper  limit for these regions we assumed $ L_{H\alpha+[NII]} = 3.8\times10^{38}\textrm{ erg s}^{-1}$, which is the integrated \ha luminosity in a 13.2 arcsec aperture assuming a constant flux per pixel equal to the r.m.s. noise of the \ha image background. The derived \ha attenuations for the complete catalogue are presented in Table~\ref{table:attenuation}. We then derived UV--corrected magnitudes  and calculated the FUV--NUV colour for each  region. These results are also presented in Table~\ref{table:attenuation}: Column 1 presents the ID of the regions; Column 2 \& 3 the equatorial coordinates; Columns 4-7 the \ha, 24~\micron\, FUV and NUV luminosities respectively; Column 8 the \ha attenuation; column 9 and 10 the FUV and NUV  attenuation--corrected magnitudes; Column 11 the FUV - NUV colour derived and Column 12 the  \hii\ region ID (taken from table \ref{table:HII})  located close to each of the corresponding UV knot.  

\begin{table*}[!]
\caption{Summary of positions and photometry of the  star--forming knots. Column 1:  ID of the regions; Columns 2 \& 3: right ascension and declination respectively; Columns 4-7: decimal logarithm of the \ha, 24~\micron\, FUV and NUV luminosities respectively, these values have not been corrected for extinction; Column 8: estimated \ha attenuation in magnitudes; Columns 9 and 10:  FUV and NUV attenuation--corrected magnitudes; Column 11: the FUV--NUV colour of the star forming region. Column 4 shows luminosities corrected for a 25\% [\nii] contamination. For the regions where Column 8 has a value of 0.0, Columns 4 and 5 correspond to an upper limit of the corresponding luminosities.
\label{table:attenuation}}
\begin{center}
\begin{tabular}{llllllllllll}
\hline
\hline
ID &      R.A.      & Decl.       & $\log~L_{H\alpha}$ & $\log~L_{24\mu m}$ & $\log~L_{FUV}$   & $\log~L_{NUV}$     & A$_{H\alpha}$   & FUV      & NUV       & FUV - NUV & H$\alpha$ ID\\
   &   (J2000)      & (J2000)    & (erg s$^{-1}$)  & (erg s$^{-1}$) & (erg s$^{-1}$) & (erg s$^{-1}$) & (mag)      & (mag) & (mag) & (mag) &  \\
\hline
1   &  09:32:11.67  &  21:30:40.80  &  $\;\;$39.15	&  $\;\;$40.92  &  40.63  &  40.73  &  1.27	&  15.74  &  15.67  &  $\;$0.08   &    2\\
2   &  09:32:12.34  &  21:30:40.00  &  $\;\;$38.93	&  $\;\;$40.57  &  40.53  &  40.61  &  1.06	&  16.37  &  16.22  &  $\;$0.15   &    14,15\\
3   &  09:32:12.25  &  21:30:23.80  &  $\;\;$38.87	&  $\;\;$40.32  &  40.80  &  40.83  &  0.78	&  16.19  &  16.03  &  $\;$0.16   &    4,20,21\\
4   &  09:32:12.75  &  21:30:11.70  &  $<$38.48 	&  $<$39.41  &  40.65  &  40.58  &  0	&  17.95  &  17.68  &  $\;$0.27   &    -\\
5   &  09:32:12.23  &  21:31:05.70  &  $\;\;$38.94	&  $\;\;$40.71  &  40.74  &  40.81  &  1.29	&  15.44  &  15.45  &  -0.01  &    1\\
6   &  09:32:10.89  &  21:30:26.10  &  $\;\;$38.65	&  $\;\;$40.47  &  40.69  &  40.83  &  1.36	&  15.44  &  15.30  &  $\;$0.14   &     - \\
7   &  09:32:11.78  &  21:30:12.50  &  $\;\;$38.86	&  $\;\;$40.32  &  40.90  &  40.91  &  0.8	&  15.89  &  15.81  &  $\;$0.08   &    2,3\\
8   &  09:32:11.85  &  21:29:55.30  &  $<$38.48 	&  $<$39.53  &  40.77  &  40.80  &  0	&  17.65  &  17.13  &  $\;$0.52   &    13\\
9   &  09:32:12.23  &  21:30:05.70  &  $<$38.48 	&  $<$39.75  &  40.71  &  40.74  &  0	&  17.79  &  17.26  &  $\;$0.53   &    51\\
10  &  09:32:10.18  &  21:29:35.90  &  $<$38.48 	&  $<$39.82  &  40.79  &  40.87  &  0	&  17.59  &  16.95  &  $\;$0.64   &     - \\
11  &  09:32:10.74  &  21:29:22.90  &  $<$38.48 	&  $<$38.94  &  40.41  &  40.54  &  0	&  18.55  &  17.78  &  $\;$0.77   &     -\\
12  &  09:32:09.78  &  21:29:22.80  &  $<$38.48 	&  $<$39.53  &  40.32  &  40.41  &  0	&  18.77  &  18.10  &  $\;$0.67   &     - \\
13  &  09:32:07.59  &  21:28:55.40  &  $\;\;$38.97	&  $\;\;$40.64  &  40.52  &  40.64  &  1.11	&  16.29  &  16.08  &  $\;$0.21   &    17\\
14  &  09:32:07.16  &  21:28:46.10  &  $\;\;$38.75	&  $\;\;$40.03  &  40.43  &  40.46  &  0.6	&  17.45  &  17.21  &  $\;$0.23   &    25\\
15  &  09:32:08.77  &  21:28:46.40  &  $\;\;$39.17	&  $\;\;$40.84  &  40.64  &  40.71  &  1.1	&  16.00  &  15.91  &  $\;$0.09   &    24\\
16  &  09:32:07.98  &  21:28:40.00  &  $\;\;$38.88	&  $\;\;$40.51  &  40.54  &  40.59  &  1.06	&  16.35  &  16.29  &  $\;$0.06   &    42\\
17  &  09:32:07.81  &  21:29:38.70  &  $\;\;$38.90	&  $\;\;$40.38  &  40.09  &  40.23  &  0.83	&  17.88  &  17.49  &  $\;$0.39   &     - \\
18  &  09:32:08.27  &  21:29:47.70  &  $\;\;$38.87	&  $\;\;$40.26  &  40.05  &  40.14  &  0.71	&  18.18  &  17.86  &  $\;$0.33   &     - \\
19  &  09:32:12.03  &  21:31:33.30  &  $\;\;$38.70	&  $\;\;$40.34  &  40.54  &  40.58  &  1.07	&  16.31  &  16.31  &  $\;$0.01   &    26\\
20  &  09:32:12.12  &  21:31:22.90  &  $\;\;$38.70	&  $\;\;$40.49  &  40.64  &  40.68  &  1.32	&  15.63  &  15.72  &  -0.1   &    41\\
21  &  09:32:10.45  &  21:31:14.50  &  $\;\;$38.83	&  $\;\;$40.70  &  40.65  &  40.75  &  1.45	&  15.36  &  15.38  &  -0.03  &    46\\
22  &  09:32:11.25  &  21:31:17.50  &  $\;\;$38.78	&  $\;\;$40.39  &  40.62  &  40.74  &  1.02	&  16.20  &  15.95  &  $\;$0.25   &    23\\
23  &  09:32:09.72  &  21:30:38.80  &  $<$38.48 	&  $<$39.53  &  40.19  &  40.34  &  0	&  19.10  &  18.27  &  $\;$0.82   &     - \\
24  &  09:32:08.78  &  21:30:34.90  &  $<$38.48 	&  $<$39.84  &  40.26  &  40.27  &  0	&  18.91  &  18.44  &  $\;$0.48   &     - \\
25  &  09:32:08.80  &  21:30:25.20  &  $<$38.48 	&  $<$39.94  &  40.17  &  40.20  &  0	&  19.15  &  18.63  &  $\;$0.52   &     - \\
26  &  09:32:09.63  &  21:30:27.20  &  $<$38.48 	&  $<$40.07  &  40.12  &  40.27  &  0	&  19.28  &  18.44  &  $\;$0.84   &     - \\
27  &  09:32:08.25  &  21:30:11.70  &  $<$38.48 	&  $<$39.85  &  40.09  &  40.06  &  0	&  19.35  &  18.96  &  $\;$0.39   &     - \\
28  &  09:32:08.45  &  21:30:00.70  &  $<$38.48 	&  $<$40.06  &  40.17  &  40.20  &  0	&  19.16  &  18.61  &  $\;$0.55   &     - \\
29  &  09:32:11.27  &  21:29:34.90  &  $<$38.48 	&  $<$39.52  &  40.37  &  40.44  &  0	&  18.64  &  18.02  &  $\;$0.62   &     - \\
30  &  09:32:11.47  &  21:29:43.50  &  $<$38.48 	&  $<$39.81  &  40.39  &  40.36  &  0	&  18.59  &  18.22  &  $\;$0.37   &     - \\
31  &  09:32:07.17  &  21:29:28.20  &  $\;\;$38.83	&  $\;\;$40.24  &  39.79  &  39.94  &  0.74	&  18.77  &  18.31  &  $\;$0.46   &    6\\
32  &  09:32:09.60  &  21:28:46.50  &  $\;\;$39.16	&  $\;\;$41.03  &  40.37  &  40.44  &  1.45	&  16.06  &  16.16  &  -0.09  &    27\\
33  &  09:32:12.54  &  21:30:53.60  &  $\;\;$38.76	&  $\;\;$40.63  &  40.45  &  40.59  &  1.46	&  15.84  &  15.75  &  $\;$0.09   &    10\\
34  &  09:32:10.51  &  21:31:38.80  &  $<$38.48 	&  $<$40.07  &  40.58  &  40.64  &  0	&  18.12  &  17.54  &  $\;$0.59   &    28\\
35  &  09:32:11.05  &  21:31:37.20  &  $<$38.48 	&  $<$40.19  &  40.54  &  40.58  &  0	&  18.22  &  17.67  &  $\;$0.56   &     -\\ 
36  &  09:32:10.29  &  21:28:42.20  &  $\;\;$38.62	&  $\;\;$40.53  &  40.40  &  40.47  &  1.53	&  15.84  &  15.98  &  -0.14  &    29\\
37  &  09:32:09.74  &  21:31:49.20  &  $<$38.48 	&  $<$39.48  &  40.42  &  40.49  &  0	&  18.52  &  17.89  &  $\;$0.63   &     - \\
38  &  09:32:08.56  &  21:31:34.30  &  $\;\;$38.76	&  $\;\;$40.44  &  40.61  &  40.65  &  1.12	&  16.05  &  16.06  &  -0.01  &    30,52\\
39  &  09:32:07.94  &  21:31:36.10  &  $\;\;$38.82	&  $\;\;$40.01  &  40.59  &  40.61  &  0.5	&  17.20  &  16.97  &  $\;$0.23   &    37\\
40  &  09:32:07.72  &  21:31:25.80  &  $<$38.48 	&  $<$39.31  &  40.54  &  40.60  &  0	&  18.22  &  17.63  &  $\;$0.59   &     - \\
41  &  09:32:10.56  &  21:28:59.90  &  $\;\;$38.92	&  $\;\;$40.38  &  40.67  &  40.72  &  0.81	&  16.48  &  16.28  &  $\;$0.19   &     -\\ 
42  &  09:32:11.30  &  21:28:56.40  &  $\;\;$38.90	&  $\;\;$40.76  &  40.64  &  40.66  &  1.42	&  15.46  &  15.64  &  -0.18  &    31\\
43  &  09:32:11.61  &  21:29:06.70  &  $\;\;$38.75	&  $\;\;$40.52  &  40.47  &  40.51  &  1.27	&  16.15  &  16.20  &  -0.05  &    45\\
44  &  09:32:10.89  &  21:28:30.90  &  $<$38.48 	&  $<$40.03  &  40.52  &  40.52  &  0	&  18.28  &  17.82  &  $\;$0.47   &     - \\
45  &  09:32:11.72  &  21:28:31.70  &  $\;\;$38.74	&  $\;\;$40.16  &  40.39  &  40.36  &  0.76	&  17.26  &  17.24  &  $\;$0.01   &    32\\
46  &  09:32:10.59  &  21:28:21.00  &  $<$38.48 	&  $<$39.53  &  40.55  &  40.57  &  0	&  18.20  &  17.71  &  $\;$0.49   &     - \\
47  &  09:32:09.78  &  21:28:14.80  &  $<$38.48 	&  $<$39.74  &  40.64  &  40.60  &  0	&  17.97  &  17.61  &  $\;$0.36   &     - \\
48  &  09:32:12.68  &  21:28:37.90  &  $\;\;$39.01	&  $\;\;$40.57  &  40.54  &  40.53  &  0.94	&  16.55  &  16.59  &   -0.04  &    33\\
49  &  09:32:13.30  &  21:28:46.40  &  $\;\;$39.07	&  $\;\;$40.46  &  41.10  &  41.04  &  0.72	&  15.54  &  15.60  &  -0.06  &    34,43\\
50  &  09:32:12.83  &  21:29:38.40  &  $\;\;$38.48 	&  $\;\;$39.71  &  40.66  &  40.59  &  0	&  17.92  &  17.64  &  $\;$0.28   &     - \\
51  &  09:32:13.41  &  21:29:49.60  &  $<$38.48 	&  $<$39.63  &  40.40  &  40.33  &  0	&  18.57  &  18.29  &  $\;$0.28   &     - \\
52  &  09:32:06.38  &  21:31:15.80  &  $\;\;$38.65	&  $\;\;$39.77  &  40.42  &  40.44  &  0.43	&  17.75  &  17.47  &  $\;$0.28   &     -\\ 
53  &  09:32:13.65  &  21:28:58.00  &  $<$38.48 	&  $<$39.77  &  40.47  &  40.49  &  0	&  18.39  &  17.90  &  $\;$0.49   &     - \\
54  &  09:32:09.14  &  21:28:11.60  &  $<$38.48 	&  $<$39.98  &  40.48  &  40.47  &  0	&  18.38  &  17.95  &  $\;$0.43   &     -\\ 
55  &  09:32:08.05  &  21:28:15.20  &  $<$38.48 	&  $<$40.14  &  40.30  &  40.40  &  0	&  18.82  &  18.11  &  $\;$0.71   &    35\\
56  &  09:32:09.67  &  21:31:08.80  &  $\;\;$39.42	&  $\;\;$41.32  &  40.25  &  40.44  &  1.5	&  16.26  &  16.08  &  $\;$0.18   &    36\\
\hline
\hline
\end{tabular}
\end{center}
\end{table*}

\begin{figure}
  \includegraphics[scale = 0.5]{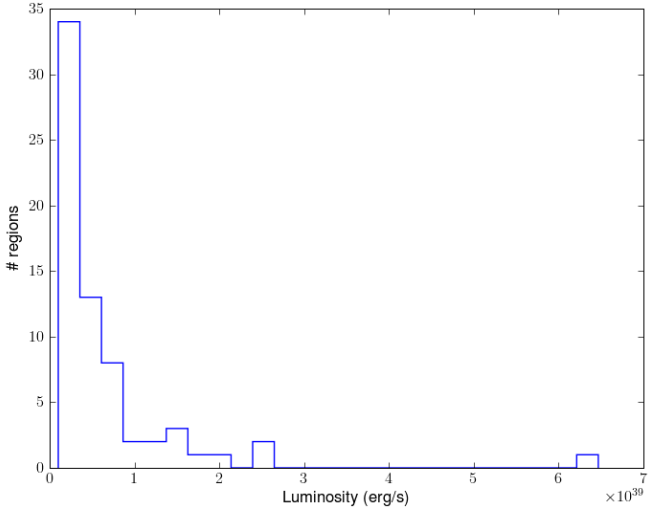}
\caption{Distribution of the \ha+[\nii] luminosity of the catalogued \hii\ regions of the bar zone of NGC~2903 from which the EW$_{H\alpha}$ were calculated.\label{fig:ha_hist}}
\end{figure}


\section{Results}
\subsection{\ha luminosity and EW$_{H\alpha}$}
\label{Result:EW}

The log of the \ha +[NII] luminosities of the \hii\ regions of NGC~2903 range  from 38.0 to 39.8 erg~s$^{-1}$ and their distribution is shown in a 
histogram  in Fig.~\ref{fig:ha_hist}. The luminosities are within typical values for extragalactic \hii\ regions in barred 
and unbarred galaxies (see e.g. Kennicutt et al. 1989; Mayya 1994; Gonz\'alez--Delgado \& P\'erez 1997\nocite{GonzalezDelgado1997}; Rozas et al. 1999\nocite{Rozas1999};  Rela\~no et al. 2005), but are higher than the \ha luminosities reported by Alonso--Herrero et al. (2001)\nocite{Alonso2001} for the nuclear region of NGC~2903. 
This apparent discrepancy is partly due to a smaller distance assumed for NGC~2903 by these authors (6.3~Mpc, versus 8.9~Mpc 
assumed in this paper). More important, Alonso--Herrero et al. (2001) apply a semiautomated method based on a limiting \ha surface 
brightness to define the edges of an \hii\ region  from an $HST$ image, yielding catalogued \ha knots which may correspond to 
parts of a single \hii\ region, as defined in ground--based  \ha imaging, and therefore have lower luminosities and sizes.

\hii\ region number 36, located in the northernmost part of the bar (see Table~\ref{table:HII} and/or Fig.~\ref{fig:HII_catalogue}), has 
an \ha luminosity of $6.4\times10^{39}$ erg~s$^{-1}$, which makes this region the  brightest \hii\ region in the galaxy (outside the nuclear region of the galaxy). Its luminosity is comparable to, for instance  the brightest \hii\ region in the nearby spiral galaxy M51 (Calzetti et al. 2005; Scoville et al. 2001)\nocite{Calzetti2005,Scoville2001} and 30 Doradus in the Large Magellanic Cloud (Kennicutt et al. 1989)\nocite{Kennicutt1989}.

The  EW$_{H\alpha+[NII]}$ of \hii~regions in the bar of NGC~2903 range from $\sim$ 71 to $\sim$ 1550 \AA, which is typical
for disk and bar \hii\ regions (e.g. Bresolin \& Kennicutt 1999; von Hippel \& Bothum 1990; Cedr\'es et al. 2005; Zurita \& P\'erez 2008)\nocite{Bresolin1997,Hippel1990,Cedres2005,zurita2008}. The distribution of EW$_{H\alpha+[NII]}$ has a mean value of  log~EW$_{H\alpha+[NII]} = $ 2.6 $\pm$ 0.3 (EW$_{H\alpha}$ in \AA). 

The \ha equivalent width of \hii\ regions depends on several factors which include the initial mass function (IMF), the evolutionary status of the \hii\ region, its metal content, and ionizing photon loss due to dust extinction or leakage from the region. 
Zurita \& P\'erez (2008) \nocite{zurita2008} analysed the effect of all these factors on the observed distribution of EW$_{H\alpha}$ for the bar \hii\ regions in NGC~1530 and we refer the reader to that paper for a more detailed discussion on the subject. 
Assuming similar physical conditions in the bar of NGC~2903, the evolutionary status of the \hii\ region is the dominant factor affecting the  EW$_{H\alpha}$ distribution. Assuming also that the \hii\ region is the product of a single burst of star formation, population synthesis models such as {\em Starburst99} (Leitherer et al. 1999)\nocite{Leitherer1999} can be used to estimate an average age for each \hii\ region, for a given IMF and metallicity.
We have therefore compared the theoretical predictions of \starburst for the evolution of a single burst of star formation with age, with our  EW$_{H\alpha}$ measurements.
Our \starburst simulations use the Geneva stellar--evolutionary tracks, and a metallicity of $Z = 0.02$ was assumed, which is the mean oxygen abundance of the gas in NGC~2903 at the range of galactocentric distances covered
by the NGC~2903 bar (Pilyugin et al. 2004)\nocite{Pilyugin2004}. For the IMF we took a multi power--law parametrization ($dN/dm \propto m^{-\alpha}$) with $\alpha = 1.35$ for $0.1 < M < 0.5M_\odot$ and $\alpha = 2.35$ (Salpeter) for higher masses up to $100 M_\odot$.
From the comparison of the \starburst output  with our measurements (after  correcting for [\nii] contamination), we estimate that the age of the catalogued \hii\ regions of NGC~2903 is within 2.7~Myr (log EW$_{H_\alpha} = 3.07$) and 7.1~Myr (log EW$_{H_\alpha} = 1.73$). Note that to derive an age from the measured equivalent width, we first had to correct the measured equivalent width for the [\nii] lines contribution.

\begin{figure}
\includegraphics[scale = 0.4]{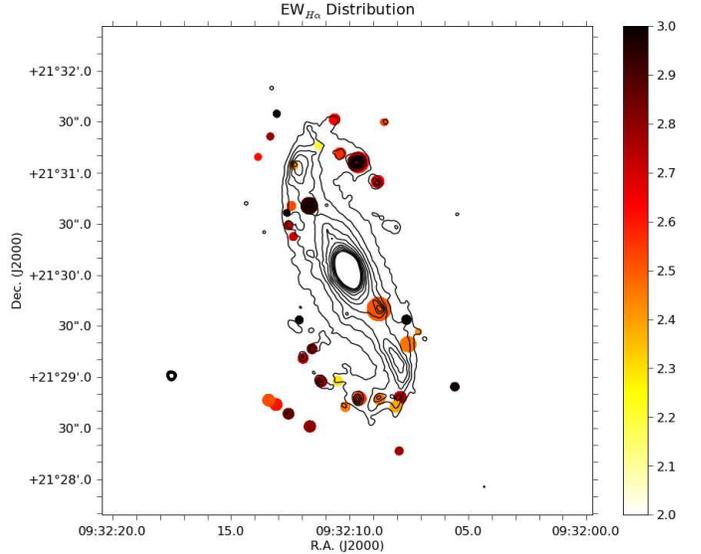}
\caption{\ha+[\nii] EW distribution in the bar region of NGC  2903. The size of the circles represents the \hii\ region projected area. The intensity contours trace the  3.6~\micron\ emission.\label{fig:EW_distribution}}
\end{figure}

Figure \ref{fig:EW_distribution} shows the distribution of the EW$_{H_\alpha}$ in the galaxy on top of 3.6 \micron\ intensity contours. As mentioned earlier, the 3.6~\micron\ emission is a strong tracer of old underlying stellar populations and therefore a good indicator of the stellar bar.
The distribution of the EW$_{H_\alpha}$ shows no correlation with the position within the bar.  

\subsection{Dust attenuation}
\label{dust_attenuation}
 
\begin{figure}
\includegraphics[scale = 0.39]{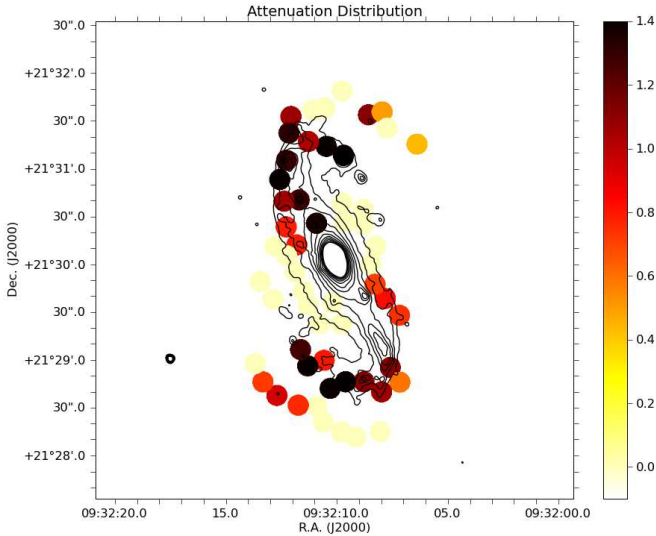}
\caption{Distribution of the \ha attenuation in NGC 2903. The circle size represents the integration apertures used  for  the UV catalogue. The intensity contours trace the  3.6~\micron\ emission.\label{fig:Att_distribution}}
\end{figure}

Figure \ref{fig:Att_distribution} presents the distribution and amplitude of the \ha attenuation in the bar of NGC~2903 with the  3.6~\micron\ intensity contours (tracing the stellar bar) overlaid on top of it. The attenuation was calculated using the ratio between the \ha emission and the 24 \micron\  (Kennicutt et al. 2007)\nocite{Kennicutt2007} as presented in Sec.~\ref{sec:dust}. The apertures are centered on regions of strong UV emission.\\
The mean attenuation value is around A$_{H\alpha} \sim 1.0$ mag and ranges from 0 to 1.5 mag (except for regions 55 and 56, closer to the nucleus with an attenuation value of 2.2). There are a few UV regions where we assume no dust attenuation due to the lack of any significant \ha and 24~\micron\ emission (see Sec.~\ref{sec:dust}). The higher attenuation values are located in the bar and at the beginning of the spiral pattern (see Table~\ref{table:attenuation}). This shows that the bar and spiral arms in NGC 2903 are gas rich, whereas the UV emitting regions to the Northwest and Southeast  of the galaxy nucleus, those regions with zero attenuation, have low gas and dust content; nevertheless,  this region is populated by young stellar clusters, see Sec.~\ref{UV_age}. Further discussion on the origin of these regions is presented in Sec.~\ref{sec:discussion}.

\subsection{Star formation rate}
\label{ha_SFR}
The locations of the most active star formation sites within a bar are determined by the favored dynamical behavior of gas and dust. Studying the distribution of the loci of current star--formation within bars  can therefore put constraints on the properties of a particular bar potential. Gas does not stay in typical bar intersecting orbits and tends to accumulate and shock in certain regions of the bar. Dust lanes can be used as tracers of  shocks in the gas flow (Athanassoula 1992)\nocite{Athanassoula1992}. From Fig.~\ref{fig:composition} one can already see that the dust lanes (as seen from the $g'-z'$ colour map) follow the CO~(J$=$1-0)  distribution, with the \ha leading the CO~(J$=$1-0)  as it has also been previously observed in other barred spirals (e.g. Sheth et al. 2002)\nocite{Sheth2002}.

The \ha emission in the bar of NGC~2903  follows an s--shape covering the whole stellar bar and it is dominated by classical localized \hii\ regions (Fig.~\ref{fig:HII_catalogue}).

In ionization--bounded \hii\ regions,the reddening corrected \ha luminosity scales directly with the total Lyman continuum flux emitted by the ionizing embedded star cluster. Therefore the \ha emission is a good tracer of current star formation. We have used the following calibration relating the \ha luminosity to the rate of star formation (Kennicutt 1998)\nocite{Kennicutt1998} 
$$
\mbox{{SFR}$_{H\alpha}$}\,(M_\odot \,\rm{year}^{-1}) = 7.9\times10^{-42}\,L\mbox{$_{H\alpha}$}\, (\rm{erg}\,\rm{s}^{-1}),
$$
for calculating the current SFR associated to each catalogued \hii\ region (Sect.~\ref{Ha_photometry}). The results are given in column~11 of Table~\ref{table:HII} and have been obtained from the observed \ha luminosities (column 4 of Table~\ref{table:HII}), and have been corrected for [\nii] emission lines contamination (see Sect.~\ref{ha}) and dust extinction inside the regions.
 The latter have been estimated as already described in Sect.~\ref{sec:dust} for the UV emitting knots, comparing the \ha\ and 24~\micron\ luminosities within the \hii\ region apertures. Given the different spatial resolution of the two images, we applied an aperture correction factor (Table~1 of Engelbracht  et al. 2007)\nocite{Engelbracht2007}. The SFRs obtained for our catalogued \hii\ regions ranges from $\sim$1.1 to $\sim$126 $\times 10^{-3} M_\odot \,\rm{year}^{-1}$. We have excluded regions 55 and 56 in this calculation due to the likely contamination from the nuclear region emission.

\begin{figure}
\includegraphics[scale = 0.39]{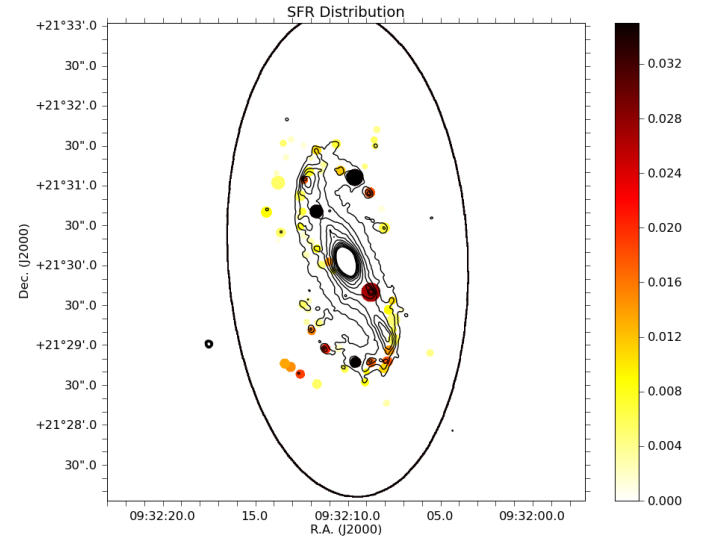}
\caption{SFR distribution in the bar region of NGC  2903. The size of the circles represents the \hii\ region projected area.  The SFR is given in $M_\odot \rm{yr}^{-1}$. The SFR has been estimated from the measured \ha luminosity of the \hii\ regions (see Sect.~\ref{ha_SFR}). The intensity contours trace the  3.6~\micron\ emission. The solid black line indicates the integration area for the calculation of the total SFR in the bar region (see Sect.~\ref{ha_SFR})}.\label{fig:SFR_distribution}
\end{figure}

The total SFR implied by the \ha  emission of the bar zone of NGC~2903 (excluding the nucleus) is $\sim 0.9 \pm 0.2\, M_\odot \,yr^{-1}$ (Fig.~\ref{fig:SFR_distribution} shows the integration area).
This value implies that the NGC~2903 bar is forming stars at a rate  comparable to other normal (non--starburst) spirals. The nucleus of NGC~2903 is almost as bright in \ha as the whole bar, implying similar SFRs (Alonso--Herrero et al. 2001; Leon et al. 2008).

Other SFR calibrations make use of the UV emission. We have derived the integrated SFR, within the same region as for the calculation of the SFR$_{H_{\alpha}}$, using the  UV emission as SF indicator. The SFR in the UV has been computed following the recipes given in Salim et al. (2007)\nocite{salim}
$$
\mbox{{SFR}$_{UV}$}\,(M_\odot \,\rm{year}^{-1})  = 1.08 \times 10^{-28}\,L\mbox{$_{FUV}$},
$$
with $L_{\rm{FUV}}$ in $\textrm{ergs s}^{-1}\rm{Hz}^{-1}$. The star formation rates calculated in this way give a SFR$_{FUV}$ = $0.44\pm 0.06$~M$_\odot$~yr$^{-1}$ for the bar region. 
The SFR calculated from both bands (cf. SFR$_{UV} = 0.44 \pm 0.06$~M$_\odot$~yr$^{-1}$ and SFR$_{H\alpha} = 0.9 \pm 0.2 $~M$_\odot$~yr$^{-1}$) and the fluxes from which these SFR have been obtained, are compatible with the results given by other works based on a large number of galaxies (i.e. Lee et al 2009; Salim et al 2007)\nocite{lee2009}.  The SFR$_{UV}$ is a tracer for {\em recent} star formation, averaged over some hundreds of Myrs (Calzetti et al 2005; Kong et al. 2004)\nocite{kong2004} while the SFR$_{H_\alpha}$ probes the {\em current} star formation.  The SFRs derived in this paper are compatible with a number of star formation histories, as derived from the models ({\em Starburst99}). It  is, however,  interesting to date the age of the UV complexes with no \ha counterpart to shed some light into the origin of these structures. This will be investigated in the following section.


\begin{figure}[t!]	
 \includegraphics[scale = 0.39]{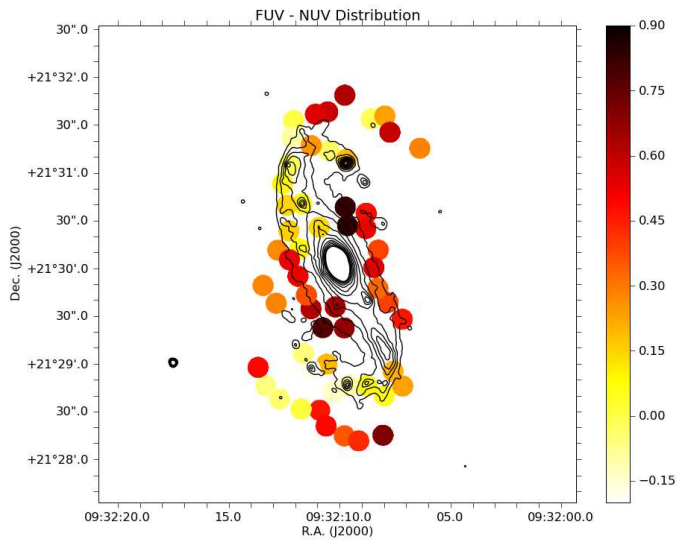}
\caption{\label{fig:UV_distribution}Distribution of the UV colours in the NGC  2903 bar and its surrounding regions. The intensity contours trace the  3.6~\micron\ emission. A concentration of redder regions is clearly visible to the Northeast and Southwest  of the nucleus. These regions correspond to the UV regions with no significant \ha counterpart.}
\end{figure}

\begin{figure}
 \includegraphics[scale = 0.4]{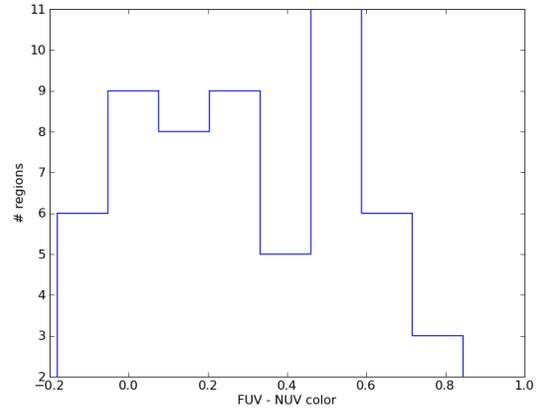}
\caption{{\em FUV - NUV} colour distribution  for the catalogued UV regions in the bar zone of NGC~2903.\label{fig:UV_hist}}
\end{figure}

\begin{figure}[!t]
 \includegraphics[scale = 0.65]{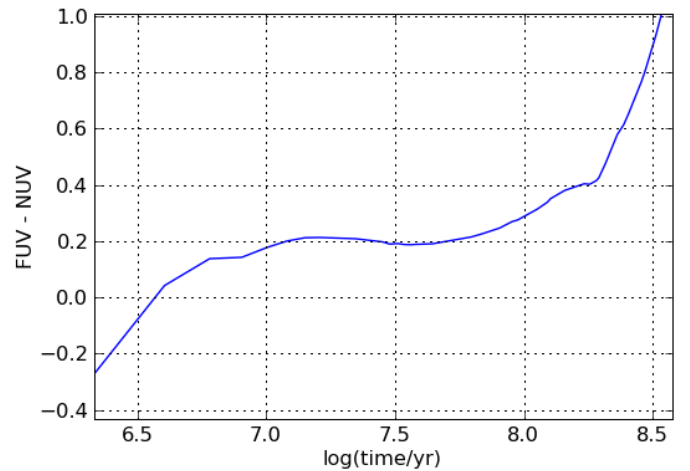}
\caption{Time evolution of the UV colour for a single starburst region. The model assumes Padova stellar tracks with Z$=$0.02 (solar) and an IMF constructed by a multi power--law parametrization with $\alpha = 1.3$ for $0.1 < M < 0.5M_\odot$ and $\alpha = 2.35$ for higher masses up to $100 M_\odot$.\label{fig:UVcolours}}
\end{figure}

\subsection{UV colours and cluster ages}
\label{UV_age}
There is now compiling observational evidence showing that the {\em FUV - NUV}  colours are very sensitive to the age of the young stellar populations due the rapid evolution of the most massive stars (i.e. Bianchi et al.  2006; Calzetti et al. 2005).\nocite{bianchi2006}
To estimate the ages of our catalogued UV regions we use the {\em FUV - NUV}  colours.  {\em FUV - NUV}  colours were calculated using fluxes corrected for dust attenuation (Sect.~\ref{sec:dust}). Figure~\ref{fig:UV_hist} shows the range of UV colours in the bar of NGC~2903. The colours have a mean value of 0.3 mag  with a standard deviation of 0.27 mag. This range of values is typical for spiral galaxies (Thilker et al. 2005; Koribalski \& L\'opez--S\'anchez 2009)\nocite{Thilker2005,Koribalski2009}.
Figure~\ref{fig:UV_distribution} shows the spatial distribution of the derived {\em FUV - NUV}  regions. One of the most striking facts on the colour distribution are the redder regions located to the Northwest and Southeast  of the galaxy nucleus and at the beginning of the spiral arm. These regions correspond mostly to those with the lowest attenuation. The regions located on the bar  consist mostly of blue UV emitting regions.

As well as age--dependent, the {\em FUV - NUV} colours are also strong tracers of metallicity; however, we can assume an approximately constant metallicity in the bar region of NGC~2903 (Pilyugin et al. 2004). With this assumption the {\em FUV - NUV} colour is, as we already mentioned, a good tracer of age. To compare our results with the theoretical models  in order to derive the ages of the clusters, we model the evolution of a single star formation burst using the \starburst population synthesis model (Leitherer et al. 1999)\nocite{Leitherer1999}. As input values we used the Padova AGB stellar tracks with a metallicity of $Z = 0.02$ (solar). For the IMF we took a multi power--law parametrization with $\alpha = 1.3$ for $0.1 < M < 0.5M_\odot$ and $\alpha = 2.35$ for higher masses up to $100 M_\odot$. We then derived {\em FUV - NUV}  colours out of the simulated spectrum at multiple time-steps from 1 Myr up to 1 Gyr (see Fig.~\ref{fig:UVcolours}).\\
In order to estimate the UV region ages, we interpolated the modelled {\em FUV - NUV} colour {\em versus} age dependence at the observed UV colours.
The colours of the bluer regions range from -0.18 to 0.15 implying an age of 3 to 10 Myr. The redder regions have colours ranging from 0.4 to 0.85 which imply ages  ranging from $\sim$ 150 to 320 Myr. The latter group of colours corresponds to regions with no significant \ha nor 24~\micron\ emission. The {\em FUV - NUV}  colour remains constant around 0.2 implying ages between 10 and a little less than 100 Myr.

\section{Discussion}
\label{sec:discussion}

NGC~2903 is a clear example of a galaxy with very different morphologies as shown by different broad--band images. Although its bar is classified in the NIR with a bar strength of 3 (Laurikainen \& Salo 2002)\nocite{Laurikainen2002}, corresponding to a relatively strong bar, and it is considered a {\it classical} bar, its morphology at shorter wavelengths is more similar to that of a patchy spiral disk.
This clearly shows that to understand the SF history in bars it is necessary to have a panchromatic view.


To investigate the properties of SF  in the NGC~2903 bar, we have at our disposal  information on the distribution of gas in two different phases, namely the ionised gas and the colder CO~($J=$1-0). As for the stellar component, we are able to trace young stars up to 1~Gyr with the UV information, the location of massive stars can be traced from the location of \hii\ regions, whereas intermediate and older stellar populations can be probed through their distribution in 3.6 \micron\ and  optical bands. 

We observe that the  CO~($J=$1-0) distribution follows closely the emission in the  3.6 \micron\ band.
There is a misalignment between the major axis of the bar as derived from ellipse fitting of optical images (Sheth et al. 2002) and the 3.6 \micron\  major axis light distribution. So, both the CO~($J=$1-0) and the 3.6\micron\  would be on the leading side of the optical photometric axis of the bar. 
The  \ha emission not only leads both the CO~(J$=$1-0)  and the 3.6 \micron\ emission but has a slightly different 
distribution, following a spiral pattern not so clearly seen in the CO~(J$=$1-0)  and 3.6 \micron\ emission. 
Also, the \ha emission leads the dust lanes as traced by optical colour maps. Numerical models predict  the gaseous component (both ionised and neutral gas) to lead the stellar bar (Martin \& Friedli  1997). 

%

This study started as a follow--up of the analysis carried out on the bar of NGC~1530 (Zurita \& P\'erez 2008) for understanding the physical processes determining star formation in bars, where it was clearly shown the importance of the bar dynamics. CO~(J$=$1-0)  and dust is
concentrated on regions where shocks and gas accumulation is favoured by the bar potential.
The distribution of the dust  in the bar of NGC~1530 is much narrower than the distribution in NGC~2903. In NGC~1530, the dust lanes are straight and very well delineated as opposed to the broken feathery morphology of the dust lanes in NGC~2903 (Fig.~\ref{fig:composition}e). 
This fact might be indicating  that  no strong shocks are present in the bar of NGC~2903, contrary to the large velocity gradients perpendicular to the bar in NGC~1530 (Zurita et al. 2004)\nocite{zurita2004}, suggesting that shocks are driving the observed morphology in NGC~1530.

Streaming motions along the bar of NGC~2903 were detected using HCN(1-0) as the kinematic tracer by Leon et al. (2008).  An undergoing analysis of the velocity gradients in the ionised gas of NGC~2903, using the data published by Hernandez et al. (2005)\nocite{hernandez2005}, will clarify whether strong velocity gradients are present or not in the bar of NGC~2903. 
Therefore, although the NIR emission suggest that NGC~2903 hosts a strong bar, the broken and wide structure of the dust lanes might be an evidence of the opposite. 

The presence of \hii\ regions in both the trailing and leading sides of the dust lanes in NGC~1530 and the evidence of an age gradient between both sets of \hii\ regions, gave support to the hypothesis (Sheth et al. 2002) that massive stars form in the trailing side of the bar dust lane and move  to the leading side. 
There are no \hii\ regions on the trailing side of the bar of NGC~2903, and no trend in \hii\ region age is found across the bar.
This may indicate that the explanation for where star formation is favoured, is  not as straightforward as in NGC~1530, or at least it is not as clearly driven by the strong bar dynamics.

The complexity of the SF properties in the bar of NGC~2903 is also shown by the  presence of compact UV sources without significant 
24 \micron\ and \ha emission. These regions are located symmetric to the center of the galaxy and nearly perpendicular to the bar as well 
as in an spiral pattern. All of these regions have FUV-NUV colours compatible with ages of a couple of hundred Myr. The lack of \ha emission 
suggests that these locations are currently not favoured by  the bar potential to form stars. These stars could have formed somewhere else in the bar and migrated afterwards to the current position. But then a few questions arise:  where have these stars formed?;  are the stars currently 
forming likely to populate similar locations? or, on the contrary, were they formed during a more general SF burst possibly linked to the bar formation, or a merger?
An undergoing analysis on the \hi\ and \ha velocity fields will shed some light into the origin of the UV complexes. 


The {\em current} SFR in the bar, as calculated from H$\alpha$, shows values similar to other disk galaxies although it has considerable high amount of dense cold gas, as traced by the HCN(1-0) (Leon et al. 2008).
The  {\em recent} SFR, as calculated from UV emission, is compatible with that of  the {\em current} SFR. 


It is worth noticing the offsets of a few arcseconds between the emission peaks in the 24 \micron\ and the \galex images. 24 \micron\ is a good tracer of current star formation while UV emission is a `time averaged' recent star formation indicator (e.g Calzetti et al 2005). The offsets found here (also seen in M51 by Calzetti et al 2005) show an evolutionary link between the sites where current star formation is occurring and the position of the UV counterpart. Estimations of the rotation curve and the pattern speed would be necessary to determine if the dynamical time--scales are compatible with the aging of the regions. 


\bigskip

\section{Summary and conclusions}
We have performed a detailed multi-wavelength study from UV to sub--millimeter observations on the NGC 2903 bar and its surrounding regions. 
We mapped and catalogued the \hii\ regions of the bar and measured their \ha equivalent widths. Furthermore we have obtained a catalogue of the UV emitting regions, and measured their peak location with respect to \ha. The extinction has been estimated using the \ha and 24 \micron\ emission.  
SFRs using both \ha and UV indicators have been calculated. We have estimated the age of the regions, using the EW$_{H\alpha}$ and the FUV-NUV colour together with stellar population synthesis models.
Our main results are:

\begin{itemize}
\item NGC 2903 is a morphologically complex galaxy. The near--infrared as well as the CO~($J$=1-0) band show a clear barred structure 
whereas the \ha and UV maps show a patchy spiral like structure. 

\item There are clear spiral like UV complexes with no significant H$\alpha$, 24~\micron\ and  CO~(J$=$1-0)  counterpart emission. These complexes are located Northwest and Southeast of the bar within the inner 1 arc minute radius (corresponding to $\sim$2.5 kpc).

\item The \ha emission along the bar, leads the CO. The 3.6~\micron\ and CO~(J$=$1-0)  emission trace each other, both leading the major axis of the optical light distribution.

\item The \ha luminosities and EW$_{H\alpha}$ of the bar \hii\ regions are within typical ranges for \hii\ regions in bar and unbarred spirals. The spatial distribution of the  \ha\ EWs does not correlate with any morphological feature of the bar.

\item The average dust attenuation in the bar area of NGC~2903 is A$_{H\alpha}$= 1.06, and ranges from 0 to 1.5 mag.

\item The {\em FUV-NUV} colour distribution is distributed in two regions, the  bluer regions range from -0.18  to 0.15 implying an 
age of $\sim$3 to 10 Myr. The redder regions have colours ranging  from  0.4 to  0.85 which imply ages  
ranging from 150 to 400 Myr. The latter  correspond to regions with no significant \ha nor 24~\micron\ emission. 

\item The SFRs of the bar region derived from \ha and from the UV emission are  $0.9 \pm 0.2\, M_\odot \,yr^{-1}$  and {\bf $0.4 \pm 0.1\, M_\odot \,yr^{-1}$ }  respectively. 
\end{itemize}

All these results suggest that an agent triggered a SF burst a few hundred Myrs ago. 
Interestingly, we see some stellar clusters unrelated to the current SF locations, symmetrically located nearly perpendicular to the bar (in the inner $\sim$2.5 kpc). The origin of these regions might be related to this SF burst.

In a following paper we will analyse the gas kinematics of the galaxy, with the aim of 
shedding some light on the  origin (merger vs. secular evolution) of these findings.

\begin{acknowledgements}
G. Popping acknowledges support by the Stichting Groninger Universiteitsfonds. I. P\'erez acknowledges support by  the Netherlands Organisation for Scientific Research (NWO, Veni-Grant 639.041.511). I. P\'erez \& A. Zurita
acknowledge support from  the Spanish Plan Nacional del Espacio de  Ministerio de Educaci\'on y Ciencia (via grant C-CONSOLIDER AYA 2007-67625-C02-02). I. P\'erez \& A. Zurita also thank the Junta de Andaluc\'{\i}a for support through the  FQM-108 project. We would like to thank Jorge Iglesias, M\'onica Rela\~no, Ute Lisenfeld and Herv\'e Wozniak for the fruitful discussions. We thank the referee for the very useful comments and suggestions that have greatly improved the manuscript.

\end{acknowledgements}

\bibliographystyle{natbib}
\bibliography{4183ref}

\end{document}